\documentclass[review]{elsarticle}

\usepackage{lineno,hyperref}
\usepackage{graphicx}
\usepackage{subfig}
\usepackage[linesnumbered,ruled,lined]{algorithm2e}
\usepackage{float}
\usepackage{tikz}
\usepackage{epstopdf}
\usepackage{amssymb}
\usepackage{amsmath}
\usepackage{booktabs}
\usepackage{array}
\usepackage{multirow}
\usetikzlibrary{arrows.meta}

\journal{X. X. X}

\begin{document}

\begin{frontmatter}
		
		\title{A Combined Data-driven and Physics-driven Method for Steady Heat Conduction Prediction using Deep Convolutional Neural Networks}
		\author[Hao's address]{Hao Ma}
		\ead{hao.ma@tum.de}		
		\author[Xiangyu's address]{Xiangyu Hu \corref{mycorrespondingauthor}}
		\cortext[mycorrespondingauthor]{Corresponding author. Tel.: +49 89 289 16152.}
		\ead{xiangyu.hu@tum.de}
		\author[Yuxuan's address]{Yuxuan Zhang}
		\ead{uruz7@live.com}
		\author[Nils's address]{Nils Thuerey}
		\ead{nils.thuerey@tum.de}
		\author[Hao's address]{Oskar J. Haidn}
		\ead{oskar.haidn@tum.de}
		
		\address[Hao's address]{Department of Aerospace and Geodesy, Technical University of Munich, 85748 Garching, Germany}
		\address[Xiangyu's address]{Department of Mechanical Engineering, Technical University of Munich, 85748 Garching, Germany}
		\address[Yuxuan's address]{Beijing Aerospace Propulsion Institute, 100076, Beijing, China}
		\address[Nils's address]{Technical University of Munich, 85748, Garching, Germany}
		
\begin{abstract}
With several advantages and as an alternative to predict physics field, 
machine learning methods can be classified into two distinct types: data-driven relying on training data and physics-driven using physics law. 
Choosing heat conduction problem as an example, we compared the data- and physics-driven learning process with deep Convolutional Neural Networks (CNN). 
It shows that the convergences of the error to ground truth solution and the residual of heat conduction equation exhibit remarkable differences. 
Based on this observation, we propose a combined-driven method for learning acceleration and more accurate solutions.
With a weighted loss function, reference data and physical equation are able to simultaneously drive the learning. 
Several numerical experiments are conducted to investigate the effectiveness of the combined method. 
For the data-driven based method, the introduction of physical equation not only is able to speed up the convergence, but also produces physically more consistent solutions. 
For the physics-driven based method, it is observed that the combined method is able to speed up the convergence up to 49.0\% by using a not very restrictive coarse reference.
\end{abstract}

\begin{keyword}
Machine learning \sep Physics field prediction \sep  Data driven \sep Physics driven \sep Convolutional neural networks \sep Convergence speed \sep Combined method \sep Weighted loss function
\end{keyword}

\end{frontmatter}

\clearpage

\section{Introduction}\label{sec:Introduction}

With abundant training methods and high-performance computing resources, machine learning has been applied for many scientific research fields, including computational physics for modeling\cite{yan2018data,zhang2015machine}, optimization\cite{kirchdoerfer2016data,christelis2017physics}, control\cite{kim2017data} and other critical tasks\cite{schmid2010dynamic,brunton2019machine}. 
A specific application is to predict physics field for reducing or avoiding the large computational cost of the traditional numerical (finite volume/element/difference) methods,  such as \emph{Computational Fluid Dynamics} (CFD) which solve \emph{Partial Differential Equations}(PDEs)\cite{anderson1995computational}.

In machine learning approaches, data-driven methods were widely used to train the model with a large amount of labeled training data. For physics field prediction, the data-driven methods can be roughly identified as indirect to train closure models\cite{wang2017physics,wu2018physics} and direct to obtain the solutions\cite{goodfellow2014generative,farimani2017deep}. 
While the indirect method has achieved great successes in recent\cite{duraisamy2019turbulence,ling2016reynolds}, the direct data-driven method has also exhibited the capability to capture physics characteristics and to provide accurate estimates without resorting to expensive numerical computations\cite{wu2018deep,bhatnagar2019prediction,jeong2015data,tompson2017accelerating}. 
However, since the training data sets are still generated from traditional numerical solutions\cite{thuerey2018well,milani2018machine,lee2018data}, it leads to an embarrassing loop that it does not truly solve the issue of expansive numerical computation. 
In addition, in some difficult cases, though a large number of training samples are used, the data-driven method may still not be able to obtain sufficiently accurate solutions. 
Choosing the work in Ref.\cite{thuerey2018well} as an example, independent of the number of training samples, considerable errors always manifest themselves in the inferred flow field just behind the airfoil.
Similar phenomenon also happens in Ref.\cite{farimani2017deep}. 
It requires another novel approach to eliminate this shortcoming other than simply utilizing an even larger training data set. 

In fact, the physics law which is unknown in data-driven methods could be explicitly employed in the learning process\cite{lee1990neural,lagaris1998artificial}. 
Raissi et al. introduced this idea into machine learning algorithms and named it as \emph{Physics Informed Neural Networks} (PINN)\cite{raissi2017physics}. 
By introducing PDEs into the loss function, PINN is able to predict the solution that satisfies physics law. 
The inferred solution is trained to obey the corresponding PDE and boundary conditions. 
The effectiveness of this physics-driven method has been demonstrated through a collection of physics problems\cite{raissi2017physicspart1,raissi2019physics,lu2019deepxde,sharma2018weakly}.
Compared with data-driven methods, it extricate machine learning from the dependence of training data, remarkably decrease the cost of data set generation\cite{sun2019surrogate,zhu2019physics,geneva2019modeling}. 
However, as shown in a single case training later, the cost of physics-driven method is typically more expensive than that of the data-driven method. 

In order to remedy the above mentioned shortcomings, we propose an idea that combining data- and physics-drivens together. 
To our knowledge, this is the first attempt that simultaneously utilizes training data and physics law to drive machine learning for physics field prediction.
Choosing the steady solution of heat conduction as an example, we first consider the original data-driven and physics-driven methods based on a deep CNN respectively and compare their learning progresses for a single case training.
Then, based on the comparisons, we propose a weighted loss function combining the effects of reference target and physics law (given as Laplace equation) for training the CNN to predict temperature fields.
After this, several numerical experiments are conducted and analyzed to study the improvements achieved by the combined method. 

\section{Preliminaries}\label{sec:Preliminaries}

We choose the steady solution of heat conduction whose physics law can be expressed with a second-order PDE, i.e. Laplace equation as an example.
Focus on this problem, we first describe the CNN architecture used and the original data- and physics-driven methods, especially their distinct loss functions.

\subsection{U-net Architecture for CNN}  

The CNN used here is based on a U-net architecture, which is firstly proposed for machine vision\cite{ronneberger2015u}. 
Including the input and output layer, the U-net architecture consists of 17 layers and corresponding convolutional blocks.
Each convolutional block has a similar structure: batch normalization, active function, convolutional calculation, and dropout\cite{paszke2019pytorch}.
 
\begin{figure}[h]
	\centering
	\includegraphics[width=1\textwidth]{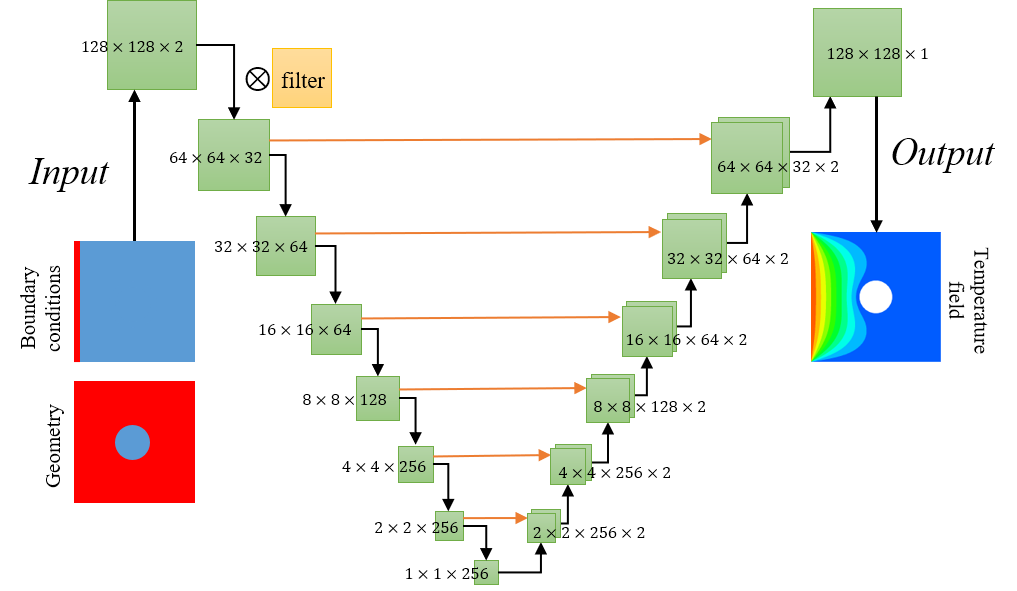}
	\caption{Schematic of U-net architecture. In the input layer, the color red and blue represent value 1 and 0 respectively. Each green box corresponds to a multi-channel feature matrix. Black corner arrows denote the down-sampling or up-sampling operation using convolutional calculation. Orange arrows denote the concatenation of the feature channels between encoding and decoding.}
	\label{figs:1-Unet}
\end{figure}

As shown in Figure \ref{figs:1-Unet}, geometry and boundary conditions are input into the architecture as square matrices with a size of $128\times128$. Then the corresponding square filters are utilized to conduct the convolutional calculation layer by layer until the matrices with only one data point are obtained. 
In this encoding process, the values of input matrices are progressively down-sampled by convolutional calculations. With the amount of feature channels increasing, large-scale information is extracted. 
Then the decoding process which can be regarded as a series of inverse convolutional operations mirrors the behavior of encoding. 
The solutions are reconstructed in the up-sampling layers along with the increase of spatial resolution and the decrease of feature channel amounts. 
Eventually the output only has one channel giving the temperature field.
It is noteworthy that there are concatenation operations between corresponding encoding and decoding blocks as shown in Figure \ref{figs:1-Unet}.
These connections effectively double the amount of feature channels in every decoding block and enable the neural networks to consider the information from the encoding layers. 

The CNNs are trained using stochastic gradient descent optimization, which requires a loss function to calculate the model error.
Except for different loss functions, the U-nets and other training settings are kept same in the following description on the original data- and physics-driven methods.
By means of backpropagation, the weights and other parameters of the entire networks are adjusted by Adam optimizer\cite{kingma2014adam} and eventually the loss is minimized and the CNN are able to reconstruct the solution of heat conduction problems.
More details of the U-net architecture and CNN can be found in Ref. \cite{thuerey2018well}.

\subsection{Original Data- and Physics-driven Methods}
 

In the data-driven method, the loss function compares the difference between training target and output result as
\begin{equation}\label{con:LossData}
	\mathcal{L}_{\rm{data}}=\left|  T_{\rm{out}}-T_{\rm{tar}}  \right|.
\end{equation}
The subscript ``data'' here denotes data-driven. $T_{\rm{out}}$ and $T_{\rm{tar}} $ are output and target temperature distributions respectively.

Based on Fourier’s law, when thermal conductivity is considered as constant and there is no inner heat source, the physics law of heat conduction can be described as two-dimensional Laplace equation
\begin{equation}
\frac{\partial^2 T}{\partial x^2}+
\frac{\partial^2 T}{\partial y^2}=
0,
\end{equation}
which is a typical PDE whose solution is important in many branches of physics. 

\begin{figure}[h]
	\centering
	\includegraphics[width=0.9\textwidth]{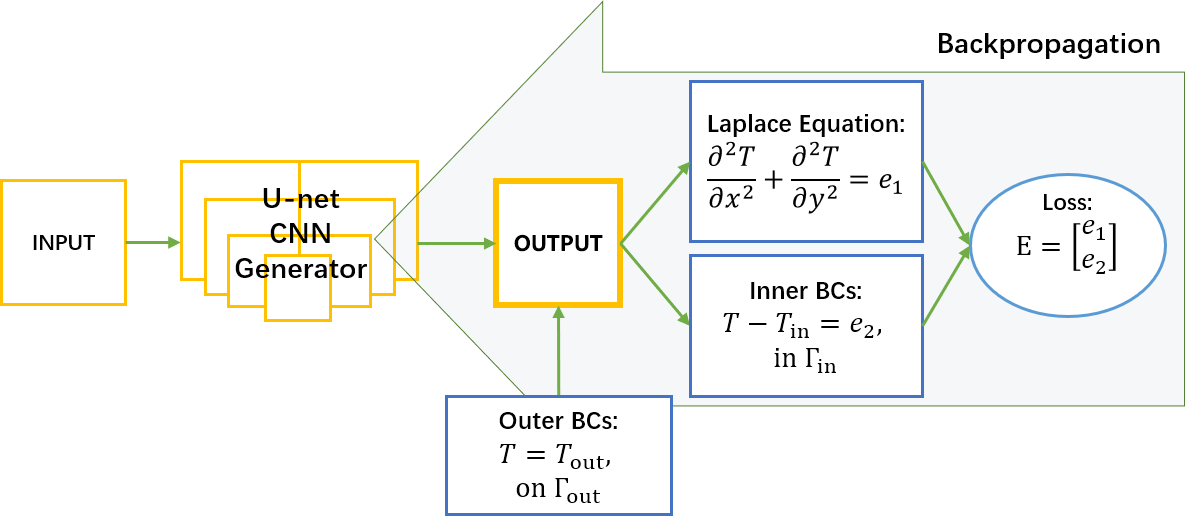}
	\caption{Schematic of the physics-driven method. U-net CNN generates the solution. The backpropagation computes the gradient of the loss function and update the weights of the multilayer CNN to satisfy the Laplace equation and boundary conditions.}
	\label{figs:2-Physics-driven Framework}
\end{figure}

In the physics-driven method, 
this physics law is used to drive the learning process by the loss function
\begin{equation}\label{con:LossPhysics}
\mathcal{L}_{\rm{phy}}=
\frac{\partial^2 T}{\partial x^2}+
\frac{\partial^2 T}{\partial y^2}=
e_{1}.
\end{equation}
The boundaries are constrained with Dirichlet boundary conditions by which the temperatures of the outer and inner boundaries are kept as a constant. 
The boundary conditions are implemented differently for the outer and inner boundaries.
While the temperatures at the outer boundaries are assigned as constants,
their values at the inner boundary as well as the inside void region 
are constrained by a loss function as
\begin{equation}
\mathcal{L}_{\rm{BC,in}}=
T-
T_{\rm{BC,in}}=
e_{2}.
\end{equation}
A schematic of the physics-driven method is shown in Figure \ref{figs:2-Physics-driven Framework}.
Note that, for the physics-driven method, the second inputting channel of the U-net CNN not only describes the geometry but also functions as a mask, 
by which the Laplace equation is not effective in the void region.

\section{Observations on Learning Processes and Combined Method}\label{sec:Observations}

In the following, we describe the observations on the learning processes of the original data- and physics-driven methods, which motivates the combined method, for a single case training.

\subsection{Single Case Training} 

\begin{figure}[!h]
	\centering
	\subfloat[Task 1]{
		\includegraphics[width=0.7\textwidth]{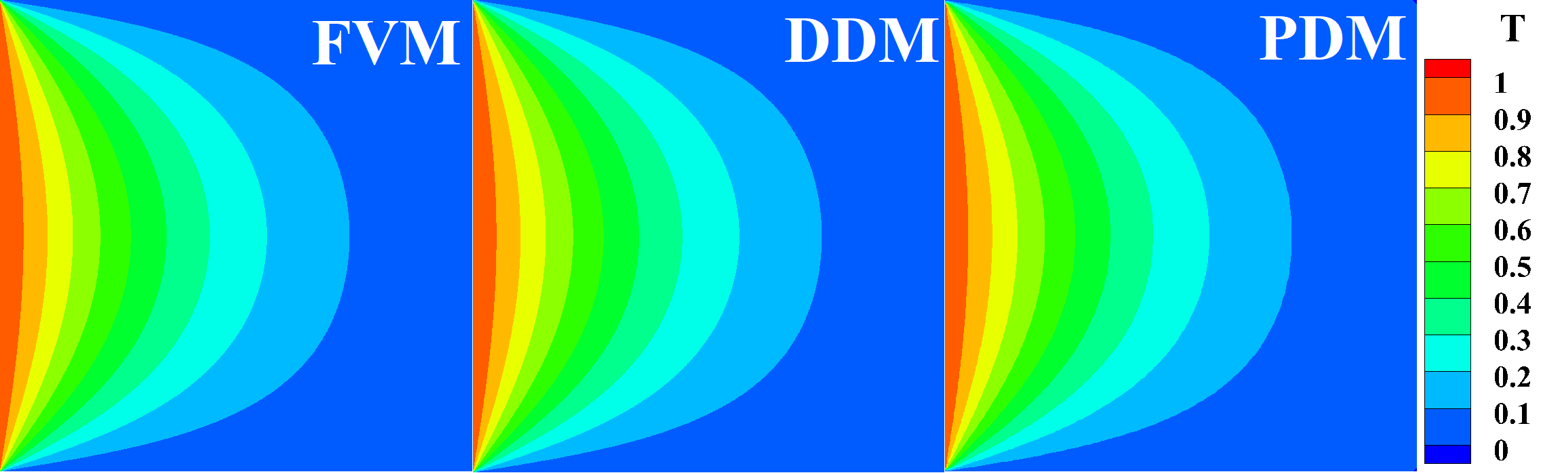}}\\
	\subfloat[Task 2]{
		\includegraphics[width=0.7\textwidth]{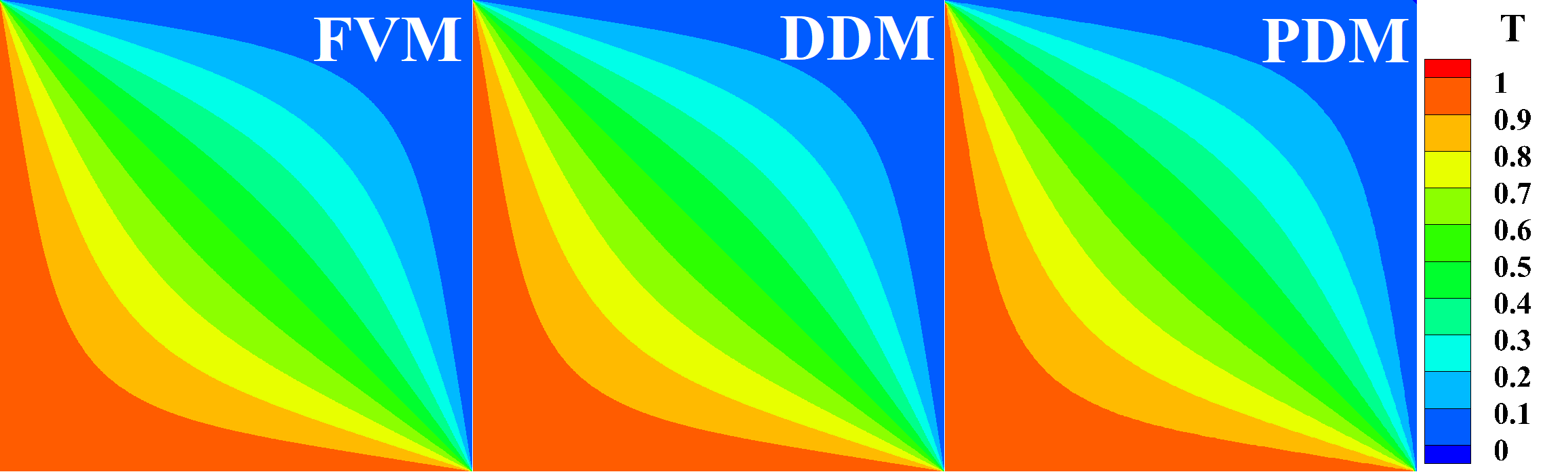}}\\
	\subfloat[Task 3]{
		\includegraphics[width=0.7\textwidth]{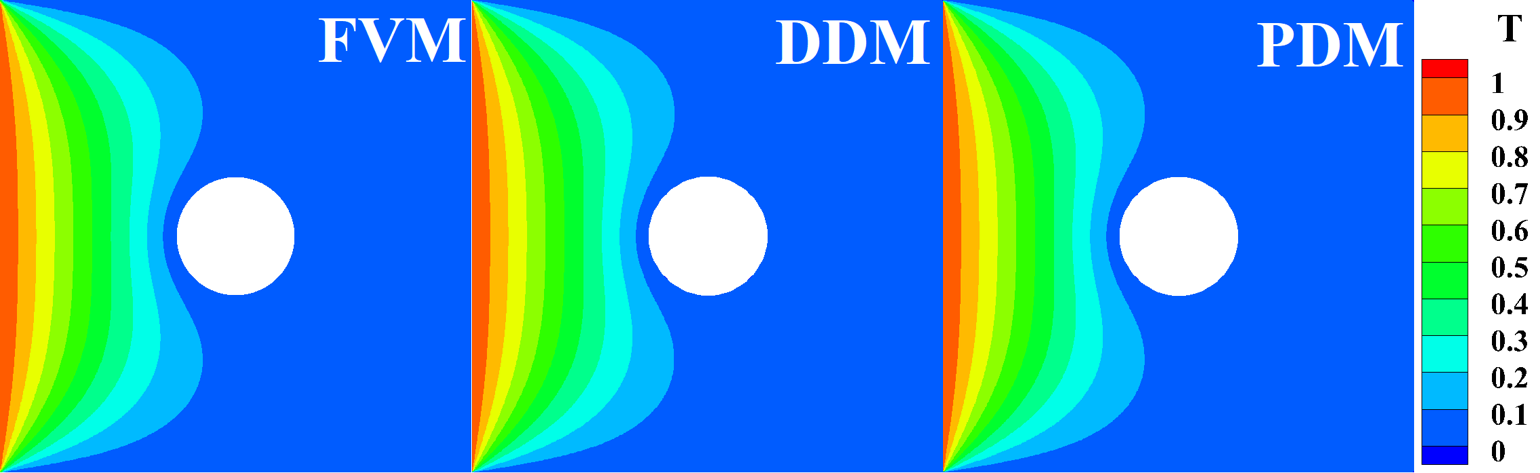}}\\
	\subfloat[Task 4]{
		\includegraphics[width=0.7\textwidth]{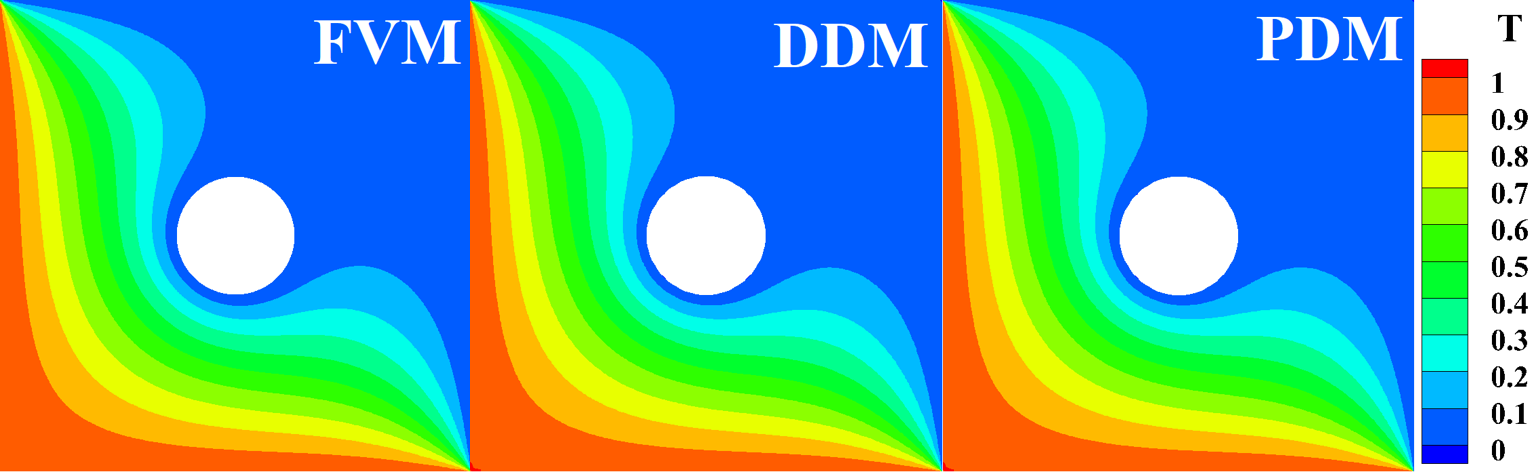}}	
	\caption{Single case training with different geometries and boundary conditions. The geometry of first two tasks is a simple square plate. 
		Task 1: the temperature of left boundary is 1, the other three are 0. 
		Task 2: the temperatures of left and bottom boundaries are 1, the other two are 0. 
		The geometry of last two tasks is a square plate with a central hole. 
		The boundary conditions of Tasks 3 and 4  are the same with Tasks 1 and 2 respectively, 
		except that the temperature of inner hole is 0. 
		Left to right: the results obtained by finite volume (FVM), data-driven (DDM) and physics-driven (PDM) methods respectively. The learned results are almost identical with numerical simulation references.}
	\label{figs:dataValidation}
\end{figure}

In general, machine learning methods are able to train multiple cases simultaneously. Here, the single case which is defined by a specific geometry and  boundary condition combination is considered and the CNN are trained to output the corresponding field solution.
By this way, 4 training tasks as shown in Figure \ref{figs:dataValidation} are carried out.  
The learning algorithms are based on PyTorch framework\cite{paszke2019pytorch} and the trainings are conducted on a single NVIDIA GeForce RTX 2080 Ti Card. 
The training samples for data-driven method are obtained by \emph{Finite Volume Method} (FVM)\cite{jasak2007openfoam} and every numerical solution is interpolated into a $128\times128$ grid to suit the learning domain with a same resolution.
Taking Task 1 as an example, it takes data-driven learning about 3k iterative steps (110 s) and physics-driven learning about 23k iterative steps (750 s) , respectively, to reach the errors less than 0.2\%. 

\subsection{Comparison of Learning Processes} 

\begin{figure}[h]
	\centering
	\setlength{\abovecaptionskip}{0 cm}
	\setlength{\belowcaptionskip}{0 cm}
	\subfloat[Data-driven learning]{
		\includegraphics[width=1\textwidth]{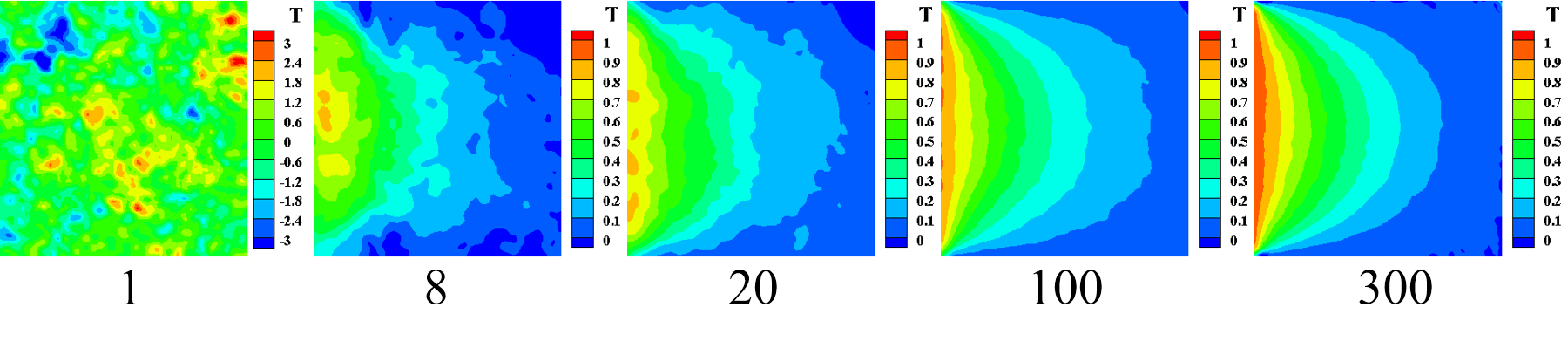}}\\
	\subfloat[Physics-driven learning]{
		\includegraphics[width=1\textwidth]{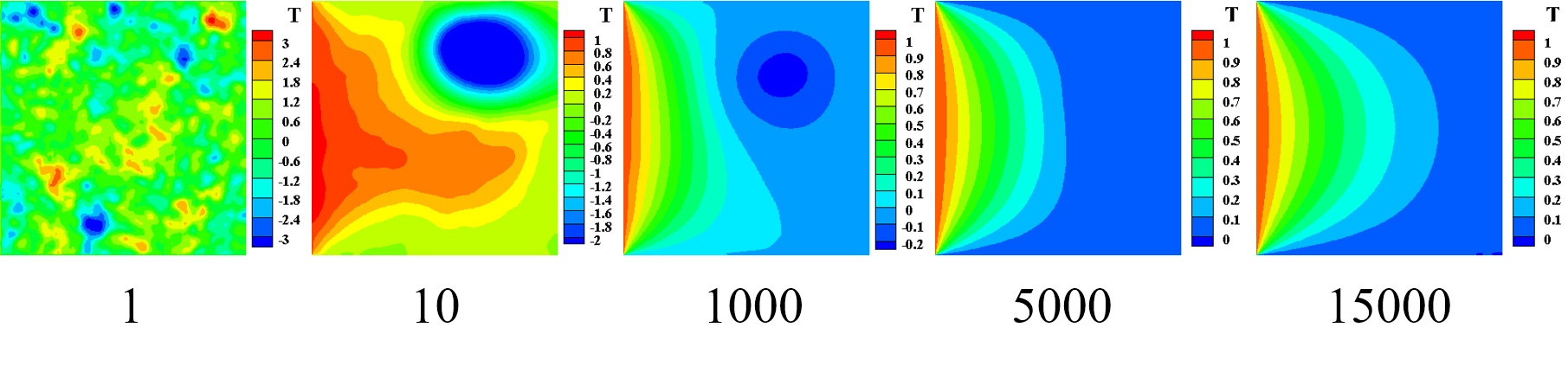}}	
	\caption{Comparison of learning process of Task 1. The numbers below the contours are iterative steps. For data-driven learning, contours transform from rough to smooth while the global structure keeps unchanged. For physics-driven learning, the contours become smooth at the early training stage and then remedy ``valley" gradually. }
	\label{figs:LearningProcess}
\end{figure}

As shown in Figure \ref{figs:LearningProcess},
in the data-driven learning process of Task 1, after a few iterative steps, a brief form of the global temperature profile (global structure) starts to appear, all local values approach those of target solution. 
The temperature contours are rough at first and smoothed successively with the global structure itself unchanged.
However, in the physics-driven learning process, the contours become smooth after a few iterative steps. Then a large error spot (denoted as a ‘valley’) appears, which also happens in the other training tasks. 
The global structure is very different from that of the true solution.
After the gradual disappearance of the ‘valley’, the still existing residuals near the boundaries make the subsequent learning process similar to a typical numerical solution process of an unsteady heat conduction problem with a Dirichlet boundary condition, as shown in the last two sub-figures of Figure \ref{figs:LearningProcess}b.

\begin{figure}[h]
	\centering
	\subfloat[Data-driven method]{
		\includegraphics[width=0.45\textwidth]{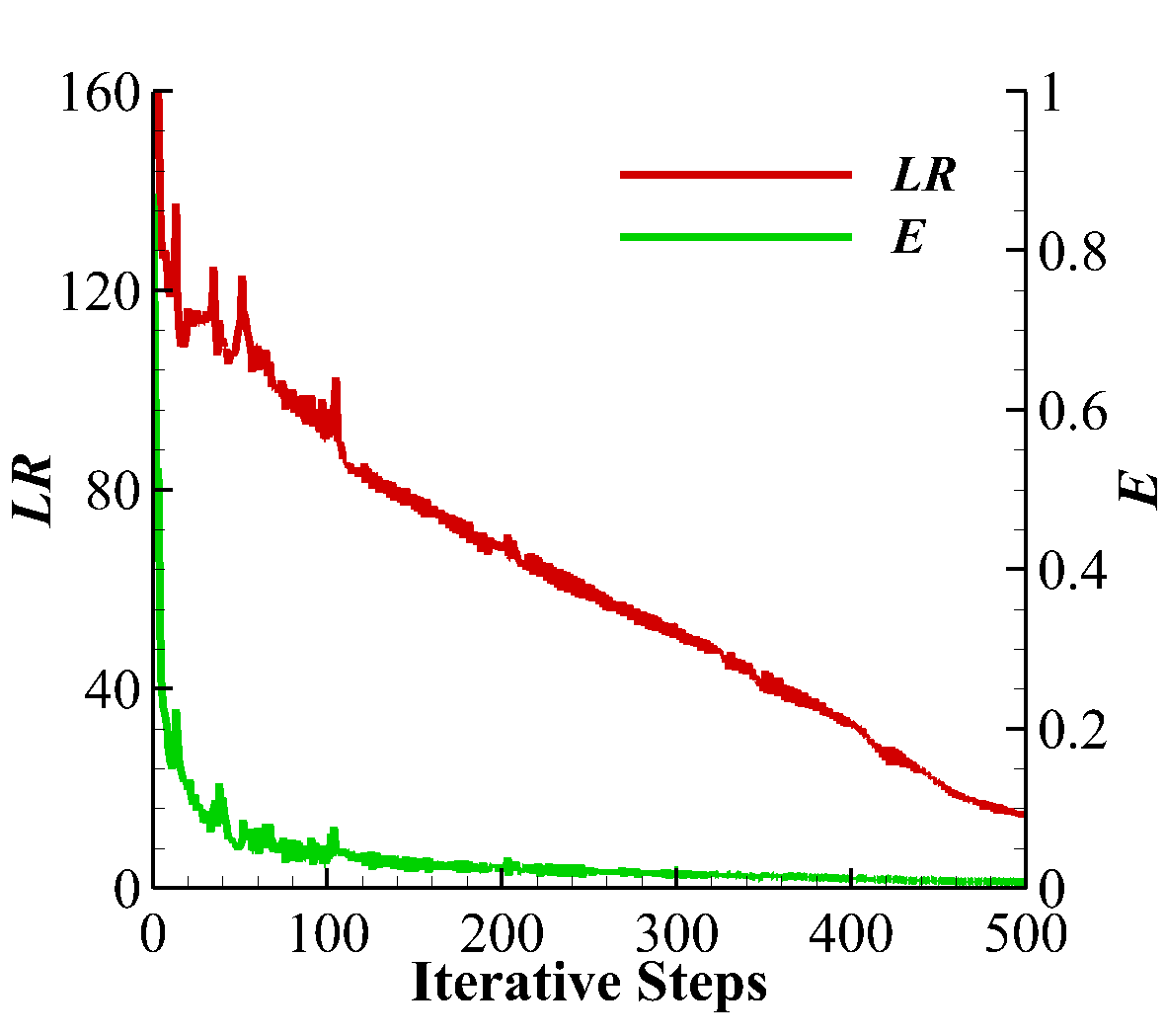}}
	\subfloat[Physics-driven method]{
		\includegraphics[width=0.45\textwidth]{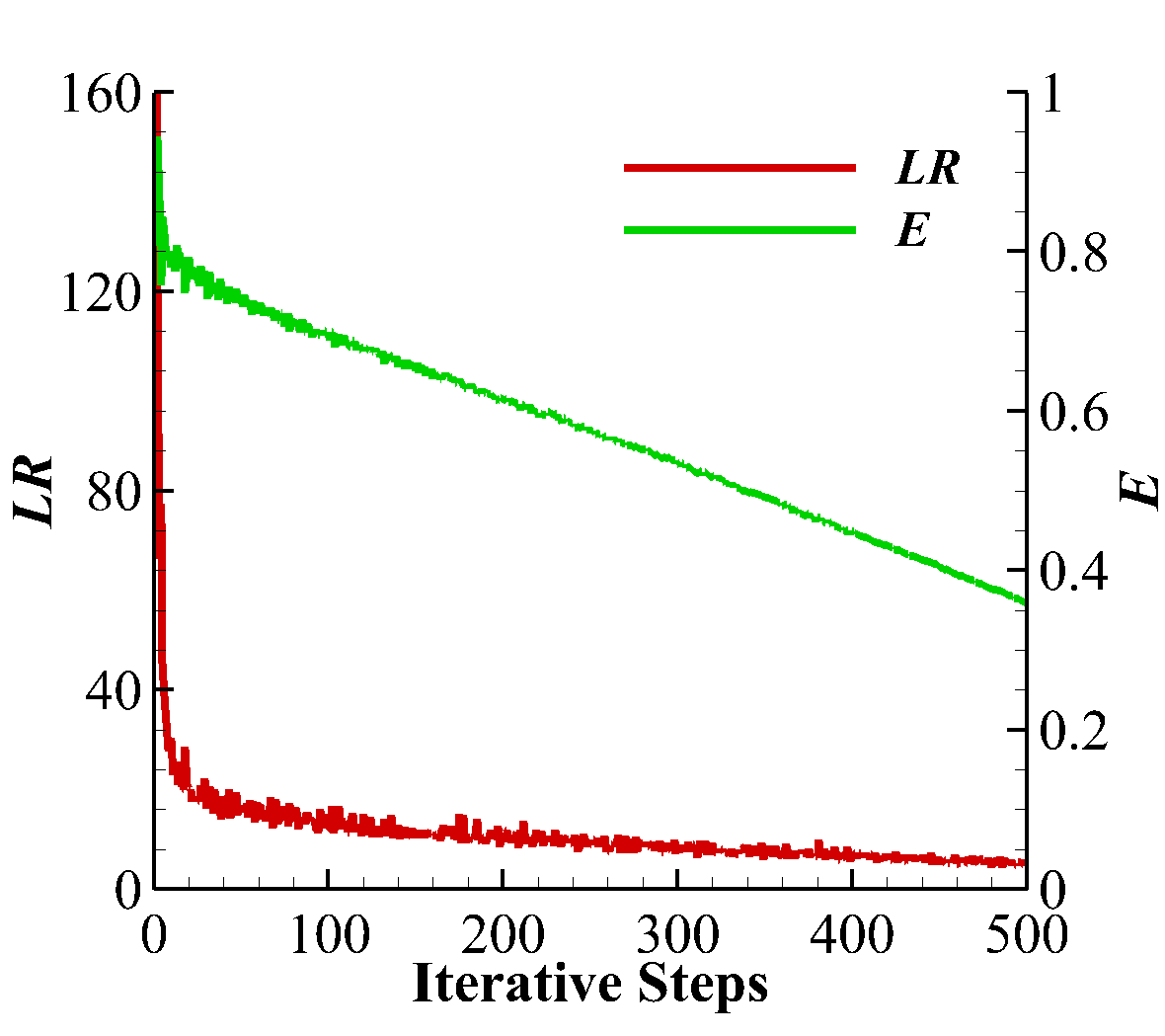}}
	\caption{Training history of $LR$ and $E$ (steps = 0 $\sim$ 500). The convergence behaviors of two learning methods have significant differences.}
	\label{figs:8-ComparisonDataPhysics}
\end{figure}

In order to study the convergence process, we defined the overall error $E$ and the Laplace residual $LR$. 
The former is defined as
\begin{equation}\label{con:E}
E=\frac{1}{n \times n}\sum_{i=1}^{n \times n} \left|  T_{\rm{out}}-T_{\rm{tar}}  \right|,
\end{equation}
where $n \times n$ is the resolution of learning domain. 
It is an estimation the degree that the outputs satisfy the solution. 
While the Laplace residual $LR$ is defined as
\begin{equation}\label{con:LR}
LR=\frac{1}{n \times n}\sum_{i=1}^{n \times n} \left|  \frac{\partial^2 T}{\partial x^2}+
\frac{\partial^2 T}{\partial y^2}  \right|.
\end{equation}
It represents the degree that the outputs satisfy Laplace equation, i.e., the local structure of solution. 
For data-driven learning, the loss function only considers the error between output and target rather than the residual of Laplace equation, so we call $E$ loss term or explicit error and $LR$ non-loss term or implicit error. Similarly in physics-driven learning, $LR$ is explicit error and $E$ implicit error.    

As shown in Figure \ref{figs:8-ComparisonDataPhysics}, for both data-driven and physics-driven learning, $E$ and $LR$ drop dramatically in the beginning.
However, after approximate 10 iterative steps, the convergence behaviors of the two terms, as explicit or implicit error, exhibits significant differences.
When the explicit errors have gradually approached an adequately small value, 
the implicit errors are still large. 
Finally, after much large number of iteration steps,
both errors decrease to sufficiently small values, 
i.e. both the solution and its local structure are obtained.

For data-driven learning, as shown in Figure \ref{figs:LearningProcess}a 
the temperature contour keeps rough for a long period. 
And there are even small isolated ‘islands’ in the neighboring temperature levels. The reason why this phenomenon happens is that the error of each data point approach to zero locally and separately. 
There is no corresponding explicit relation between these adjacent data points to restrict the local structures, which is given as Laplace equation in this case.
For physics-driven learning, 
the relation of adjacent data points or local structure is constrained by Laplace equation explicitly 
and the values varies smoothly. 
However, due to the lack of explicit restriction to target solution or the global structure, the implicit error 
$E$ decreases slowly and the convergence of the solution is much slower than that of the data-driven learning.

\subsection{Combined Method}\label{sec:Combined Method}

Based on the above observations, we propose to improve physics consistency and increase the convergence speed by combining both $E$ and $LR$ being into the loss terms.

It is observed that the scale of $LR$ is significantly larger than $E$ as shown in Figure \ref{figs:8-ComparisonDataPhysics}. 
Utilizing a simple summation of these two errors as the total loss leads a skewed optimization\cite{panchapagesan2016multi} with a dominance of the Laplace residual. 
In order to remedy this issue, we employ a weighted loss function which has been widely used in object detection\cite{redmon2016you} and audio detection\cite{phan2017dnn}. 
The weighted loss function considering both target data and Laplace equation is written as
\begin{equation}\label{con:weightedLossNew}
\mathcal{L}=
\mathcal{L}_{\rm{data}}+
{\rm R}*\mathcal{L}_{\rm{phy}},
\end{equation}
where $\rm R$ is a constant hyperparameter which is tuned to adapt the scales.  
With this weighted loss function, the different loss terms can be easily scaled to an equivalent magnitude. 
The combined method actually has two types: data-driven based and physics-driven based.
For the data-driven based method, the loss function is Equation (\ref{con:weightedLossNew}) and employed during the whole learning process.
For the physics-driven based method, the loss function is modified as
\begin{equation}\label{con:phybasedLoss}
\mathcal{L}=
\begin{cases}
\mathcal{L}_{\rm{data,ref}}+{\rm R}*\mathcal{L}_{\rm{phy}}&\mathcal{L}\geq\mathcal{L}_{\rm thr}\\
\mathcal{L}_{\rm{phy}} &\mathcal{L}<\mathcal{L}_{\rm thr}
\end{cases},
\end{equation}
where $\mathcal{L}_{\rm{data,ref}}$ is the error term with some reference targets depending on different practical considerations.
$\mathcal{L}_{\rm thr}$ is the threshold value of the loss indicating that 
once the loss is less than the threshold, the loss function will only consist of the Laplace term.   

\section{Numerical Experiments}\label{sec:Results}

\subsection{Data-driven Based Training}

The data-driven based training uses two distinct data sets: one has only a single sample and the another consists of multiple samples.

\noindent  
\emph{Single Case Training}

\begin{figure}[!h]
	\centering
	\includegraphics[width=0.55\textwidth]{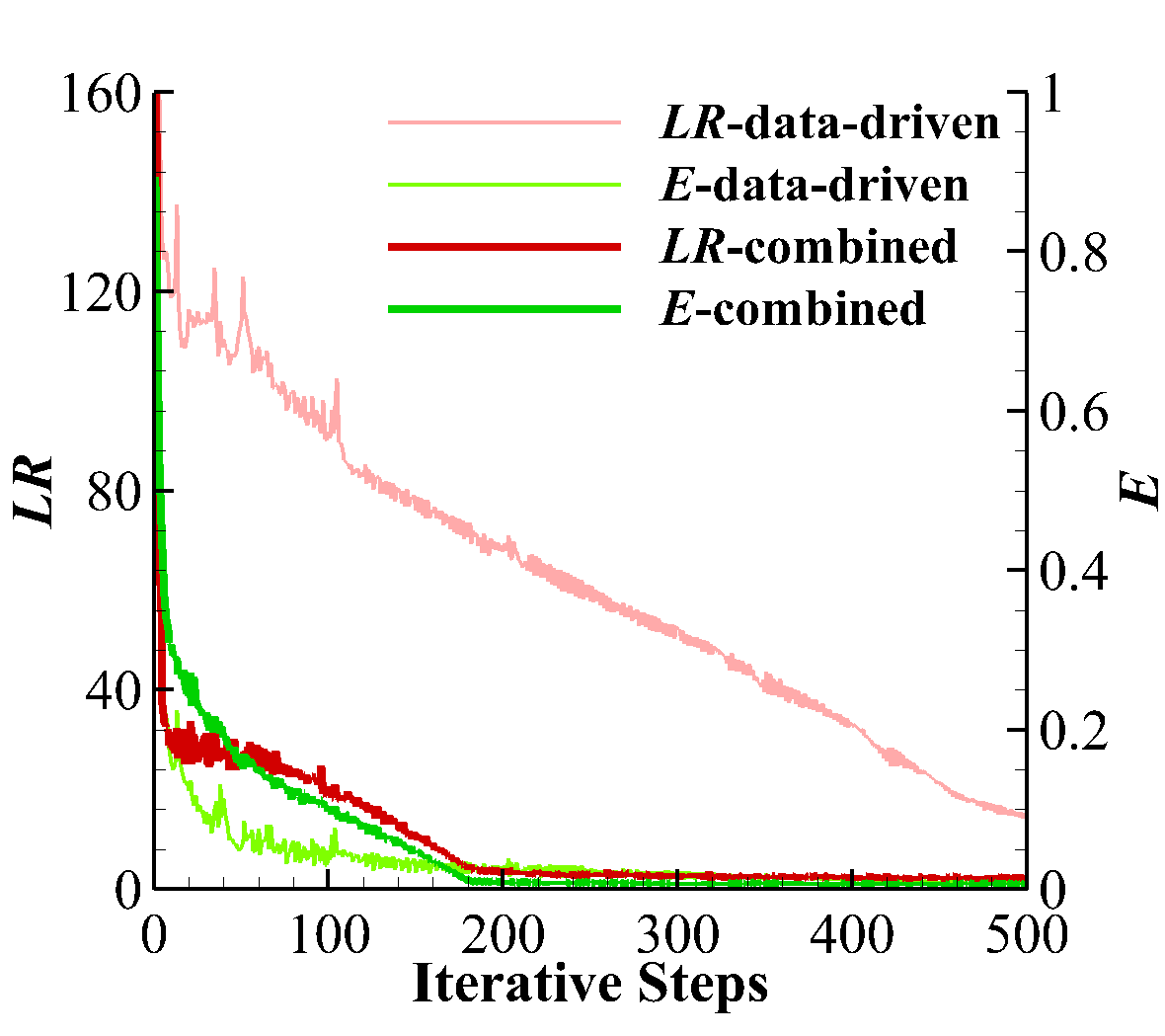}
	\caption{Enhancements for data-driven method in single case training: $LR$ and $E$. The convergence speed of $LR$ has a notable improvement.}
	\label{figs:10-ComparisonDataCom}
\end{figure}

\begin{figure}[!h]
	\centering
	\includegraphics[trim = 0cm 0.2cm 0cm 0cm,clip,width=1\textwidth]{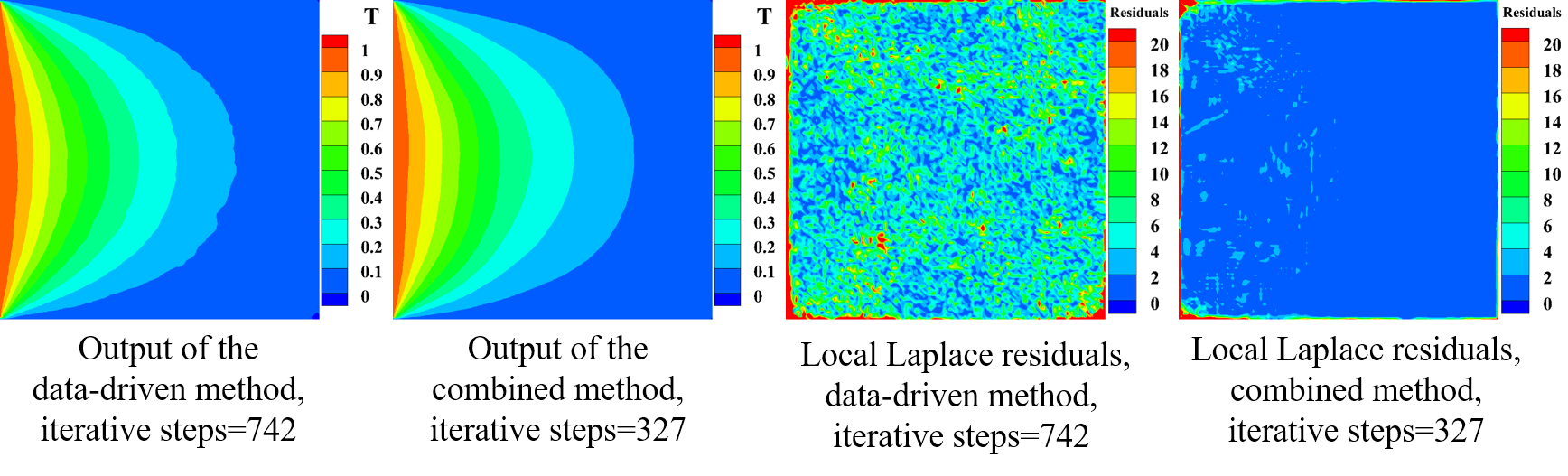}
	\caption{Enhancements for data-driven method in single case training: CNN outputs and local Laplace residuals. $E$ of the two methods are both 0.05. Compared with the data-driven method, the contour of combined method is smoother and the local Laplace residuals in data points are much smaller.}
	\label{figs:11-fieldComparison}
\end{figure}

As shown in Figure \ref{figs:10-ComparisonDataCom}, unlike the original method, $LR$ and $E$ of combined method exhibit a similar convergence behavior. After the dramatic drop in the beginning period, they change to the relatively slow decrease and then the steady decline together. $LR$ has obtained a considerable acceleration of convergence.
To check the overall performance, instead of a single run, 5 independent runs are conducted with the original and combined methods. 
It is observed that the averaged iterative steps when $E$ reaches the criteria of convergence (0.005) are 684 and 267 respectively. This result gives that the combined method remarkably accelerates the learning with a rate of 60.9\%.
In addition, the combined method also leads to a notable improvement on obtaining the physics-consistent solution, even though the overall errors have a same level (Figure \ref{figs:11-fieldComparison}). 
This is due to that the Laplace residuals are much smaller, i.e. the local structure of solution has a better physics consistency. 

\noindent 
\emph{Multiple Cases Training }

For multiple cases training, the data set is obtained with the variation of boundary temperature ($T_{boundary}=0/0.5/1$), hole shape (square/round) and position (9 different positions). There are 4,374 samples split randomly into training/test sets with an 80/20 ratio. 

\begin{figure}[!h]
	\centering
	\subfloat[Data-driven method]{
		\includegraphics[width=0.45\textwidth]{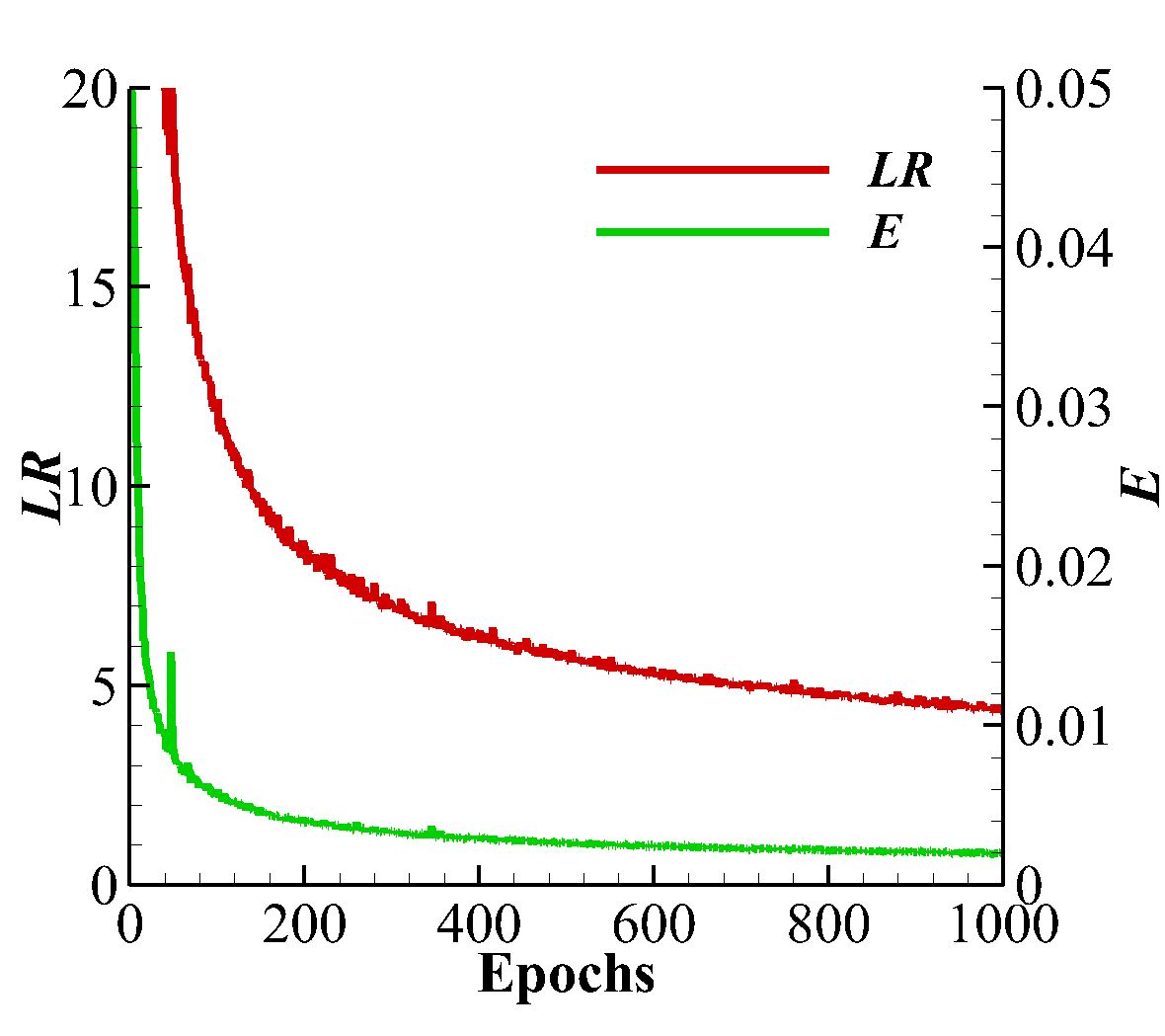}}
	\subfloat[Combined method]{
		\includegraphics[width=0.45\textwidth]{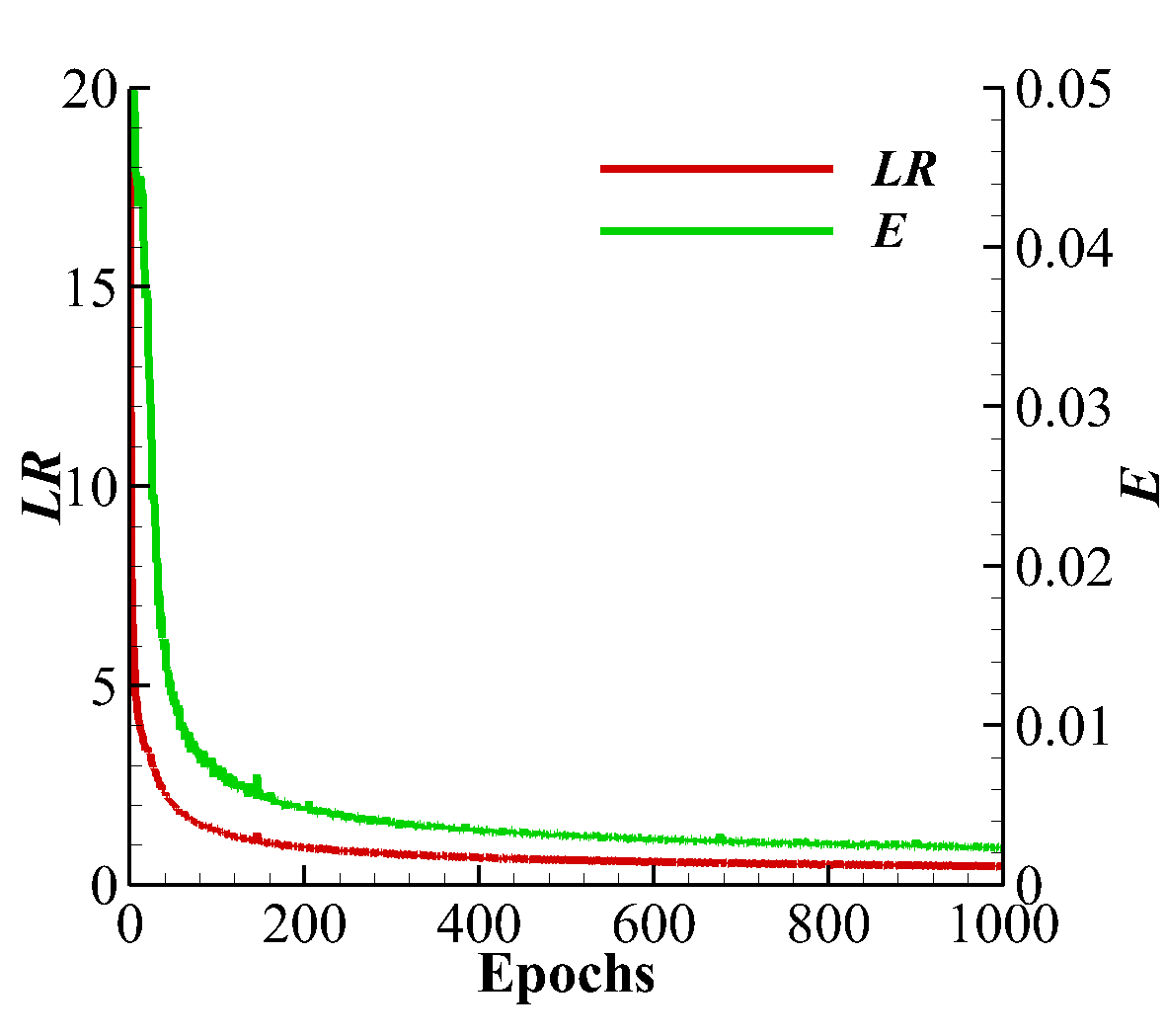}}
	\caption{Enhancements for data-driven method in multiple cases training): $LR$ and $E$. The histories of $E$ of the two methods are almost same, while the $LR$ of combined method decreases much faster.}
	\label{figs:12-ComparisonDataComMulti}
\end{figure}

\begin{figure}[!h]
	\setlength{\abovecaptionskip}{0.cm}
	\setlength{\belowcaptionskip}{0.cm}
	\centering
	\includegraphics[trim = 0cm 0cm 0cm 0cm,clip,width=1\textwidth]{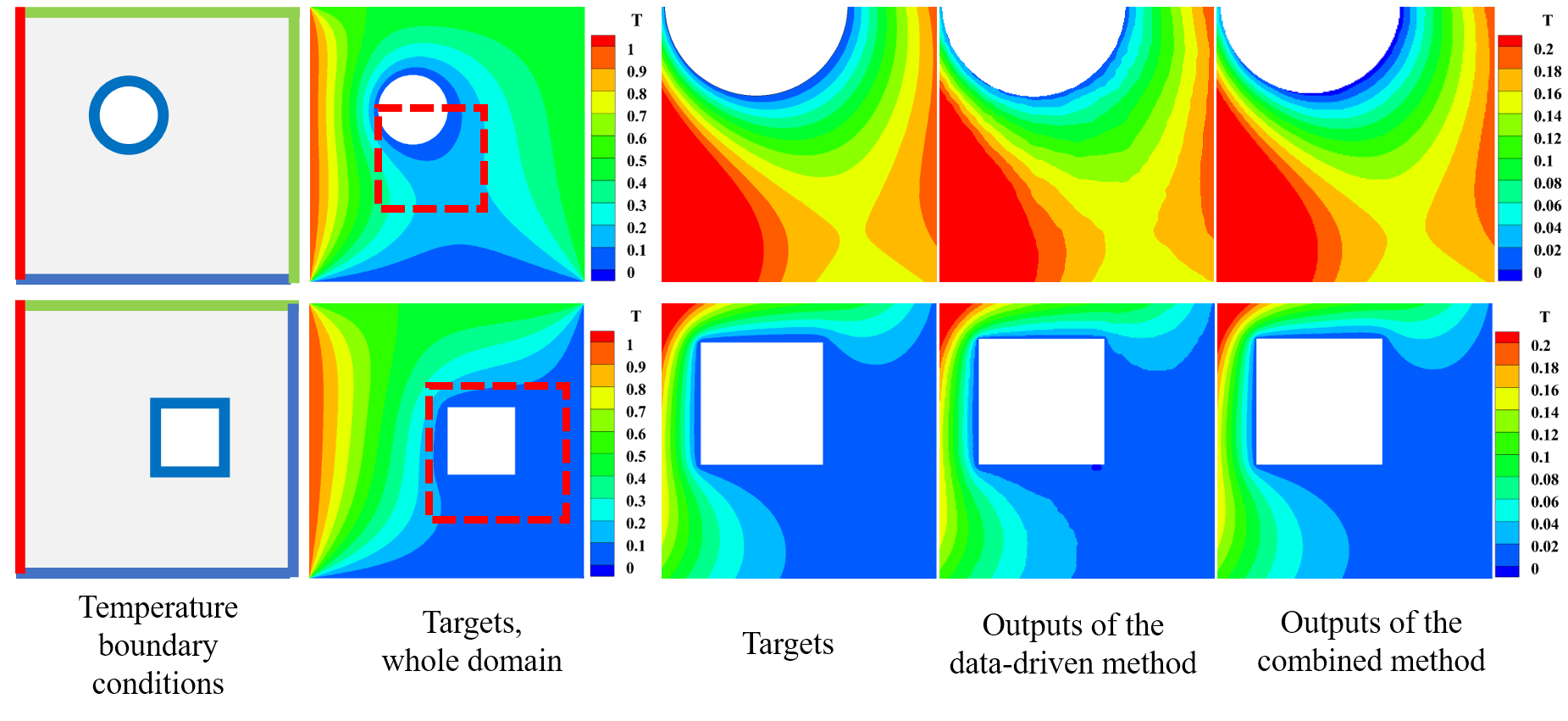}
	\caption{Enhancements for data-driven method in multiple cases training: outputs of CNN. There are two typical cases from test set with different geometries. The colors in first column represent $T_{boundary}=0/0.5/1$. The last three columns show part of the output which marked with the red dot line in the second column. In contrast to the original data-driven method, the temperature contours obtained by combined method are smoother.}
	\label{figs:13-comResults}
\end{figure}

As shown in Figure \ref{figs:12-ComparisonDataComMulti}, the decrease trend of $E$ are almost same for the original and combined methods. When the training reaches 1000 epochs, both of them are sufficiently small. However, $LR$ of them exhibit different convergence behaviors. 
The $LR$ of combined method, as loss term, decreases much faster.
For the temperature profile, the enhancement obtained by combined method is similar to that in the single case training.
From the zoomed-views (as shown in Figure \ref{figs:13-comResults}), the temperature contours obtained from combined method are much smoother, which suggests the solution is more consistent with physics law.
\subsection{Physics-driven Based Training} 

\begin{figure}[!h]
	\centering
	\subfloat[$T_{\rm true}$]{
		\includegraphics[width=0.19\textwidth]{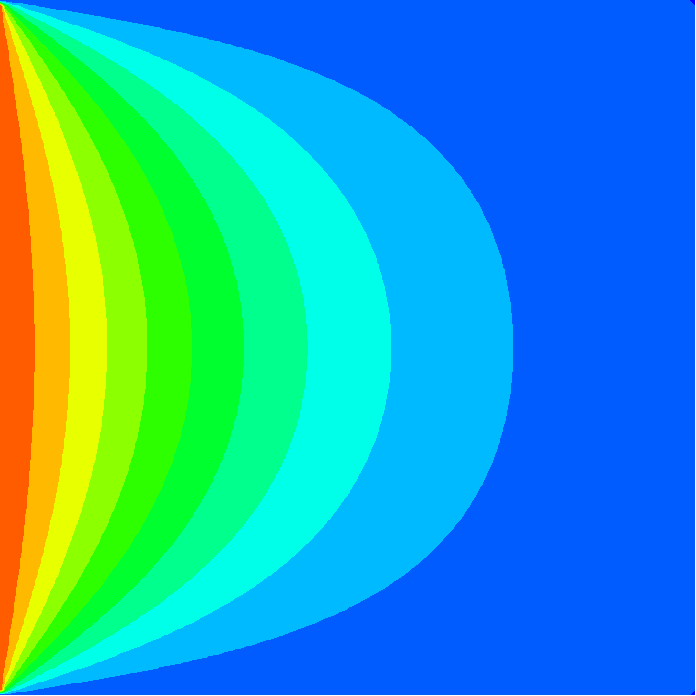}}
	\subfloat[$T_{\rm zero}$]{
		\includegraphics[width=0.19\textwidth]{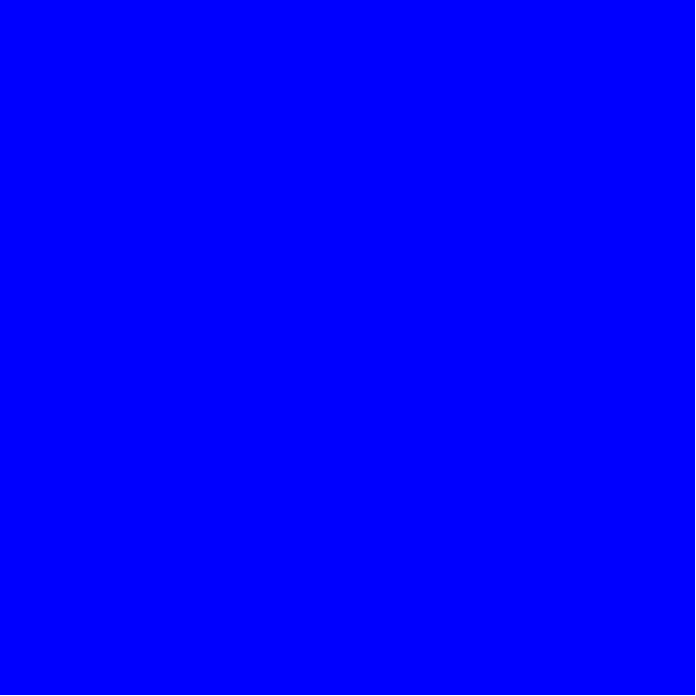}}
	\subfloat[$T_{\rm coarse}$]{
		\includegraphics[width=0.19\textwidth]{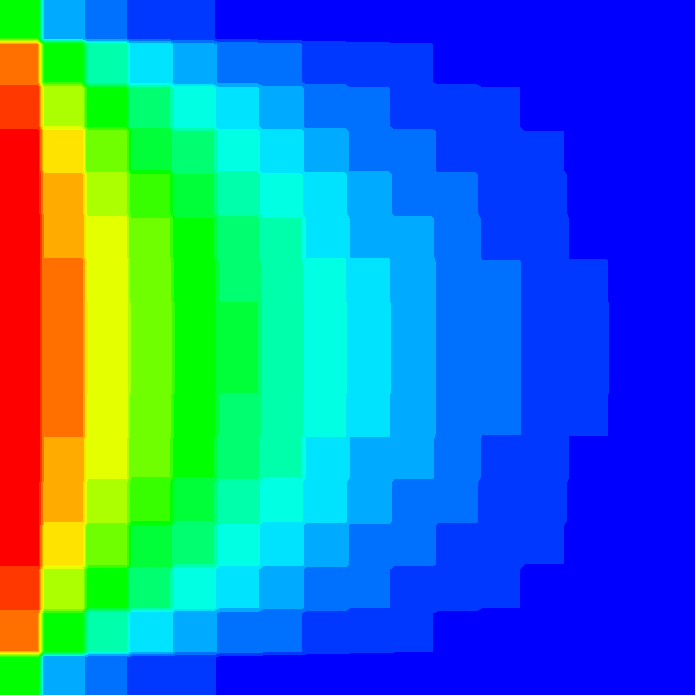}}
	\subfloat[$T_{\rm systematic}$]{
		\includegraphics[width=0.19\textwidth]{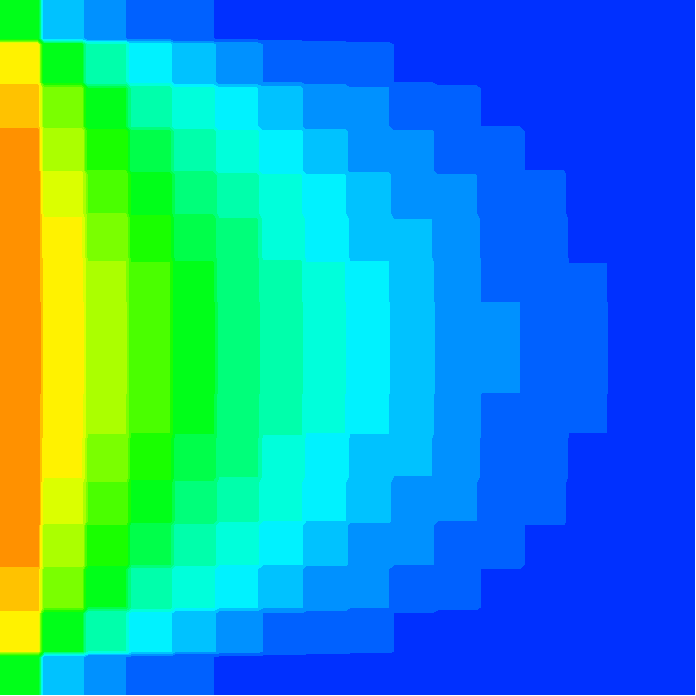}}
	\subfloat[$T_{\rm random}$]{
		\includegraphics[width=0.19\textwidth]{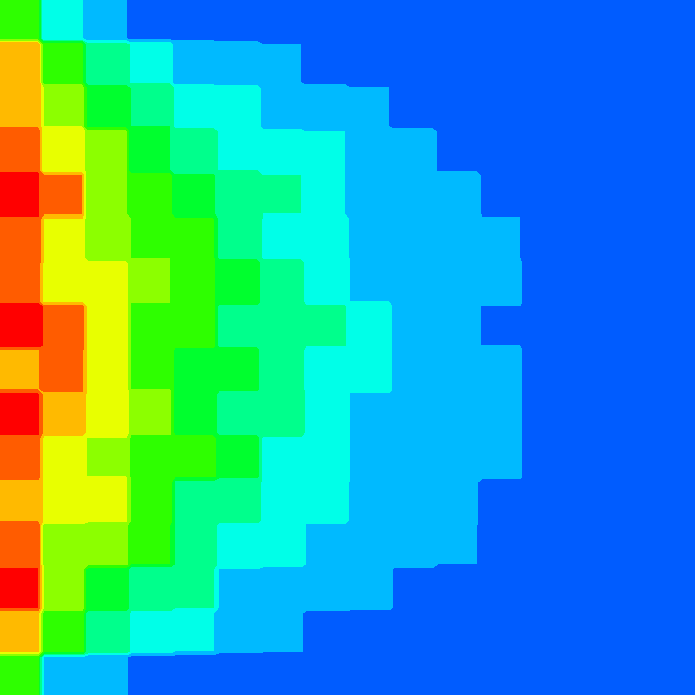}}
	\caption{Reference targets. $T_{\rm true}$ and $T_{\rm zero}$ are true and zero temperature profile respectively.
		$T_{\rm coarse}$ is down-sampled from the true temperature profile, which could be a result of numerical simulation with a coarse mesh. 
		$T_{\rm systematic}$ is the coarse profile with a different boundary temperature and represents the experimental results with systematic errors. 
		It may also estimate whether a reference target can be applied for training other cases. 
		$T_{\rm random}$ is the coarse temperature profile with random errors which represents an experimental result with measurement uncertainties.}
	\label{figs:Reference}
\end{figure}

In this section, single case is considered for physics-driven based training. According to Equation (\ref{con:phybasedLoss}), a reference target is required beforehand.
Here, as shown in Figure \ref{figs:Reference}, 5 reference targets are chosen.
Among these targets, the true and zero profiles represent the true or false limits, while the 3 coarse temperature profiles mimic the practical application where the accurate solution is not available.

\begin{figure}[!h]
	\centering
	\subfloat[Physics-driven Method]{
		\includegraphics[width=0.33\textwidth]{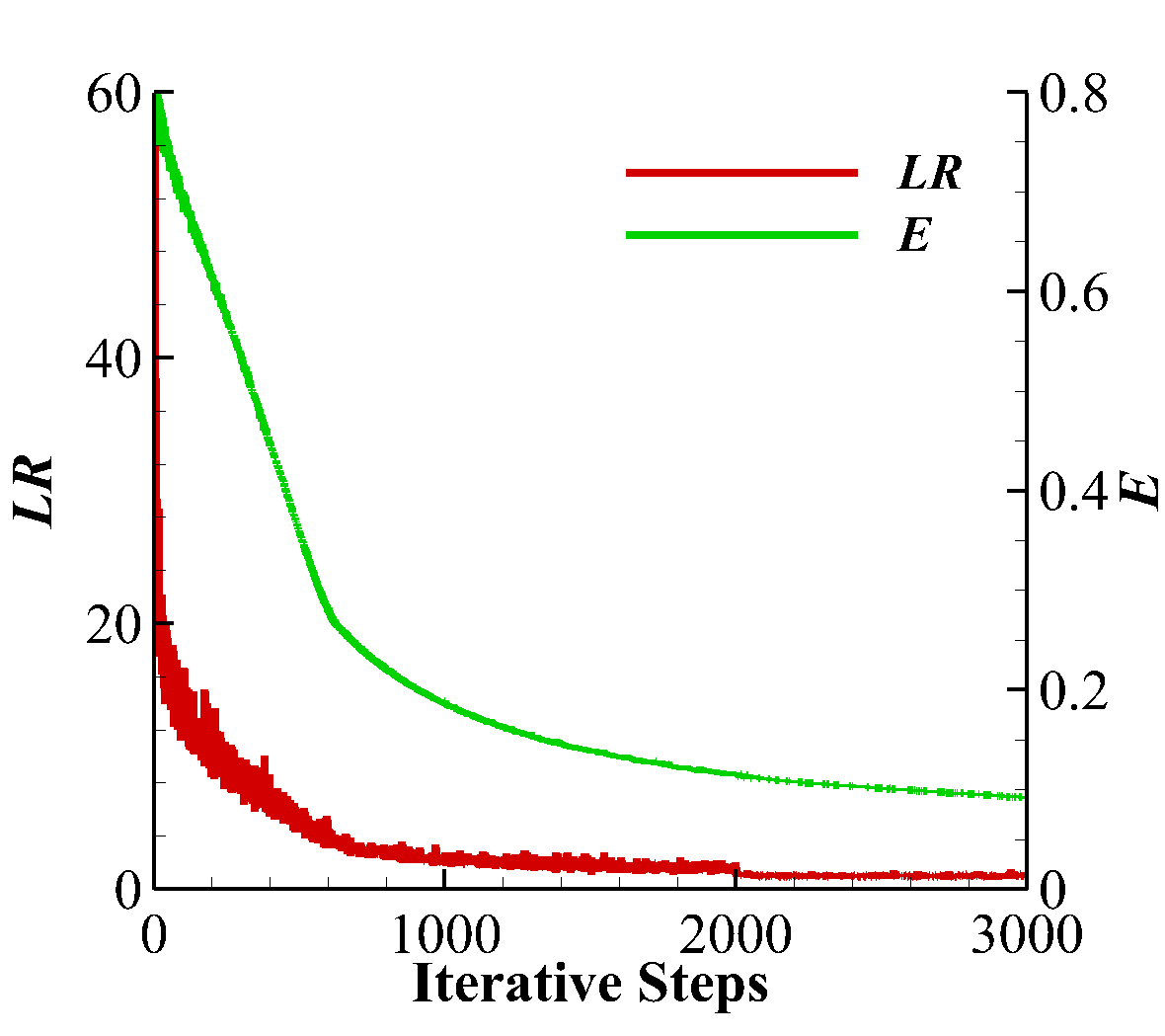}}
	\subfloat[$T_{\rm tar} =T_{\rm true} $]{
		\includegraphics[width=0.33\textwidth]{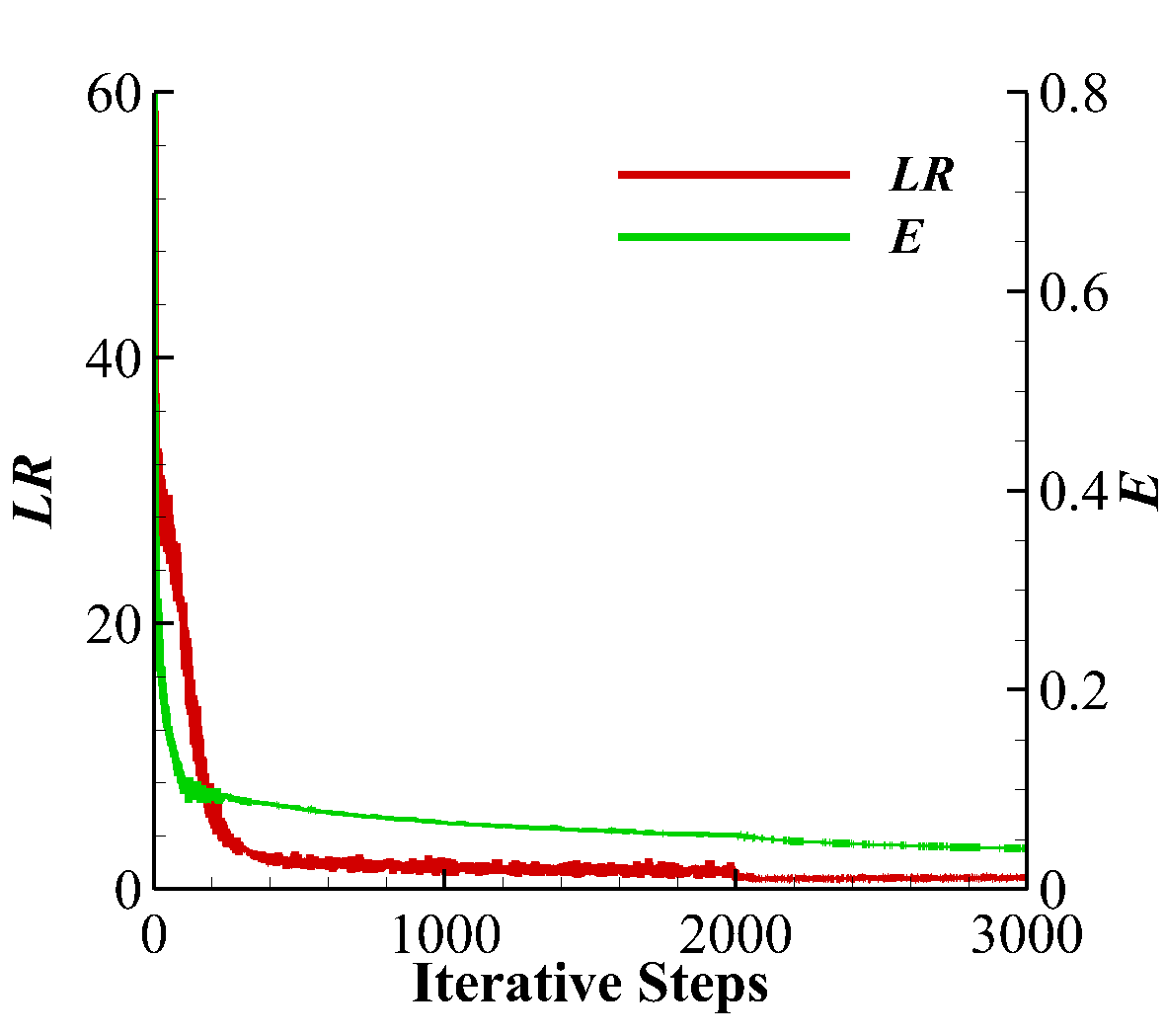}}
	\subfloat[$T_{\rm tar} =T_{\rm zero} $]{
		\includegraphics[width=0.33\textwidth]{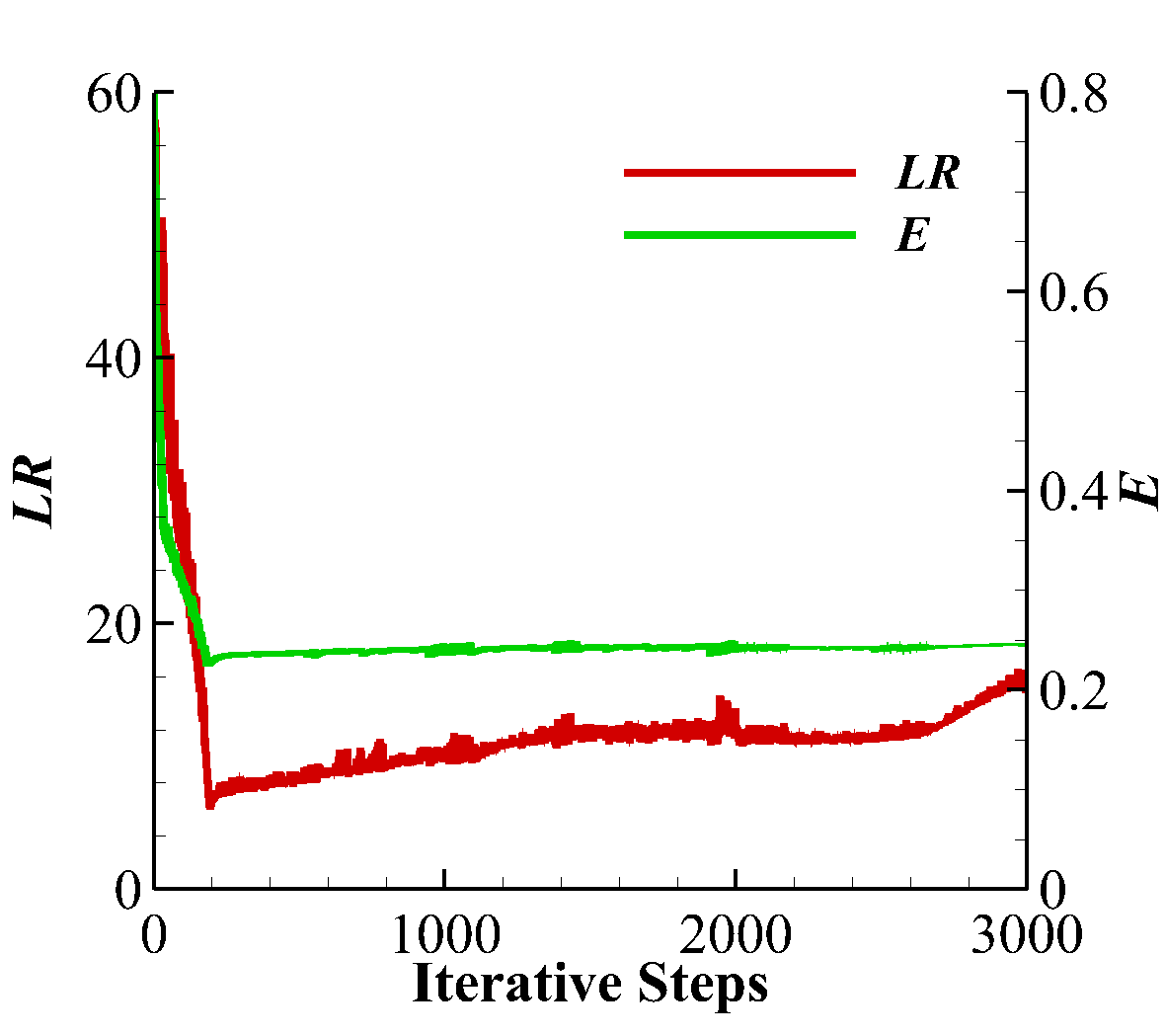}}\\
	\subfloat[$T_{\rm tar} =T_{\rm coarse} $]{
		\includegraphics[width=0.33\textwidth]{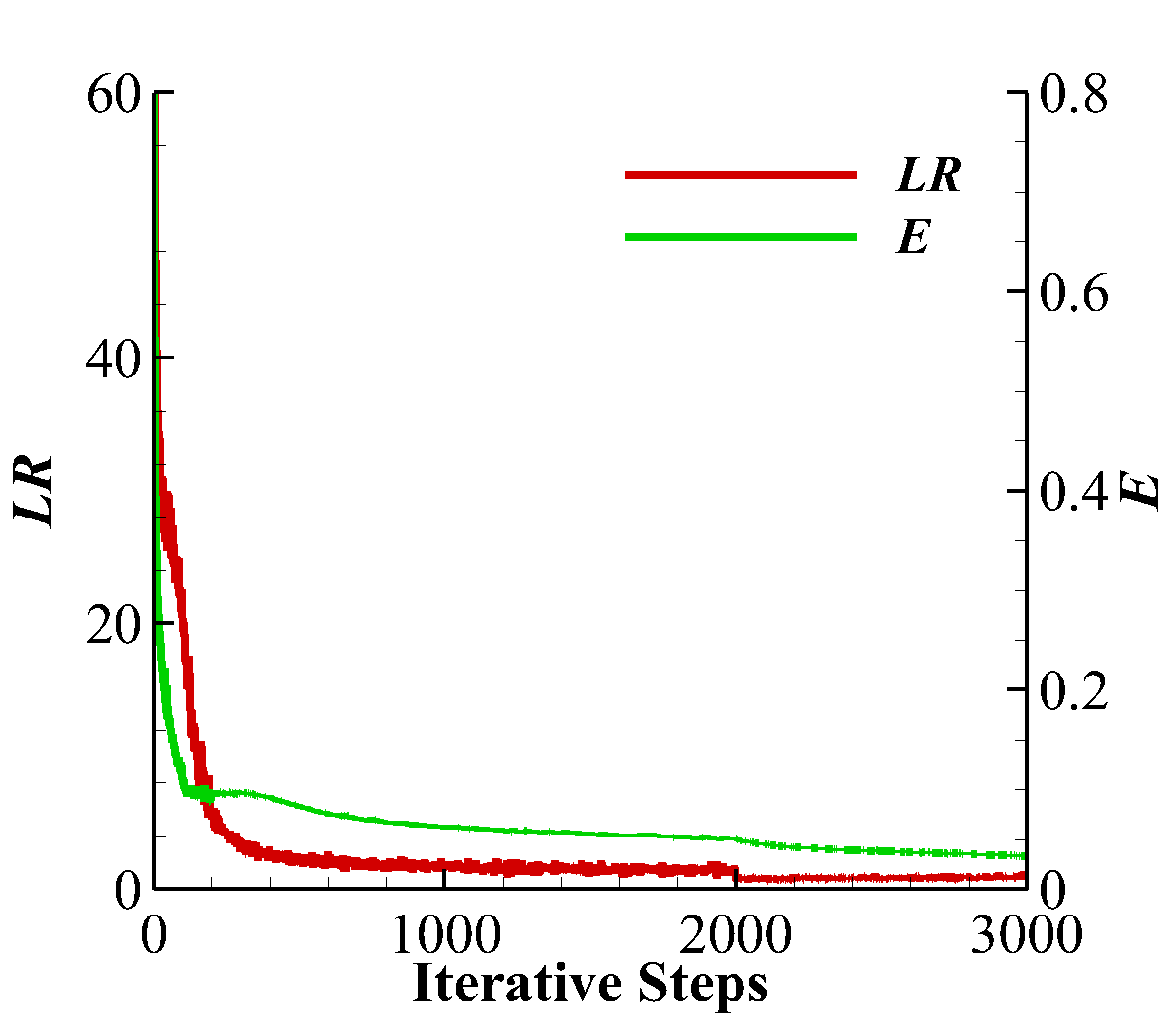}}
	\subfloat[$T_{\rm tar} =T_{\rm systemaitc} $]{
		\includegraphics[width=0.33\textwidth]{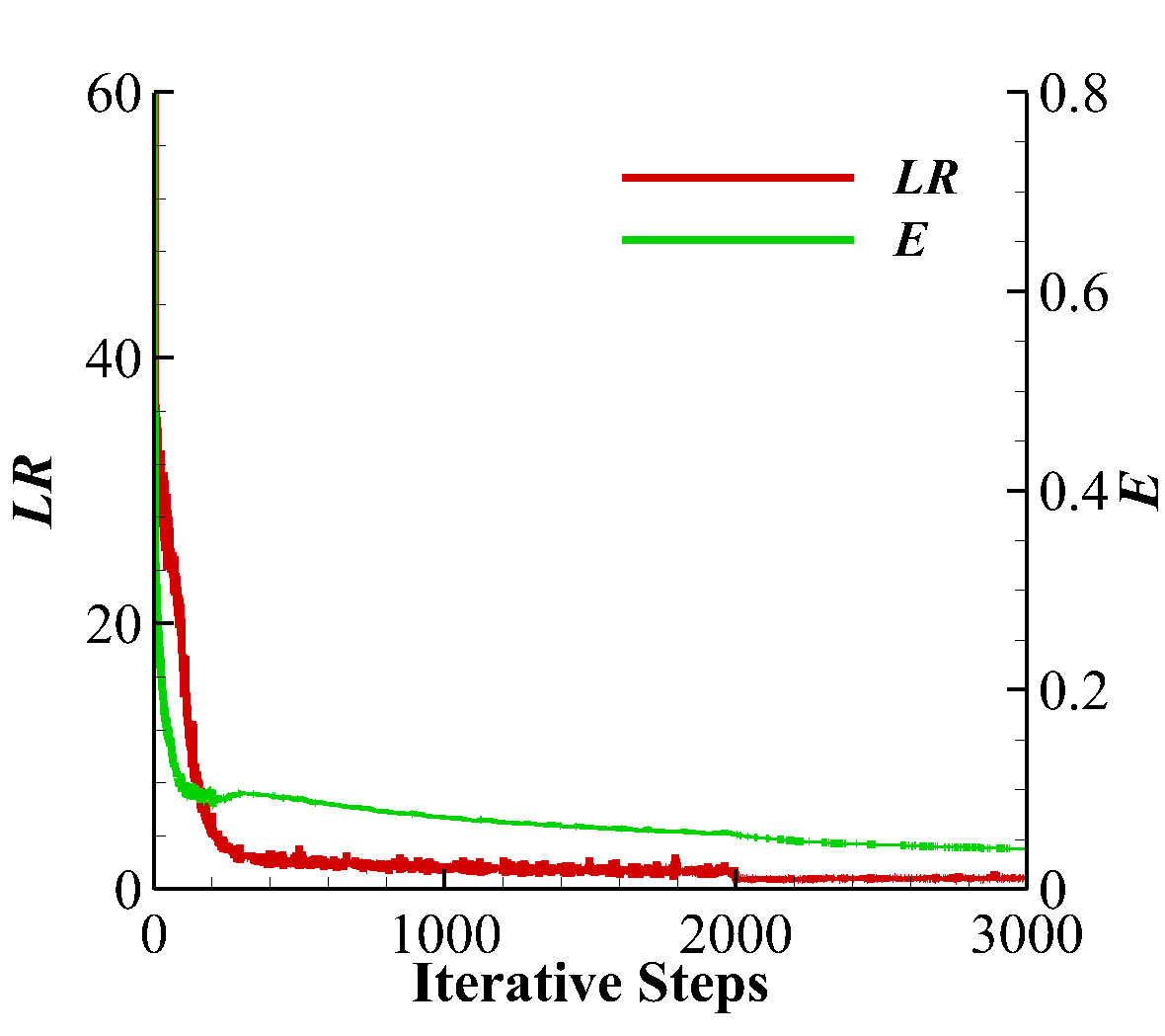}}
	\subfloat[$T_{\rm tar} =T_{\rm random} $]{
		\includegraphics[width=0.33\textwidth]{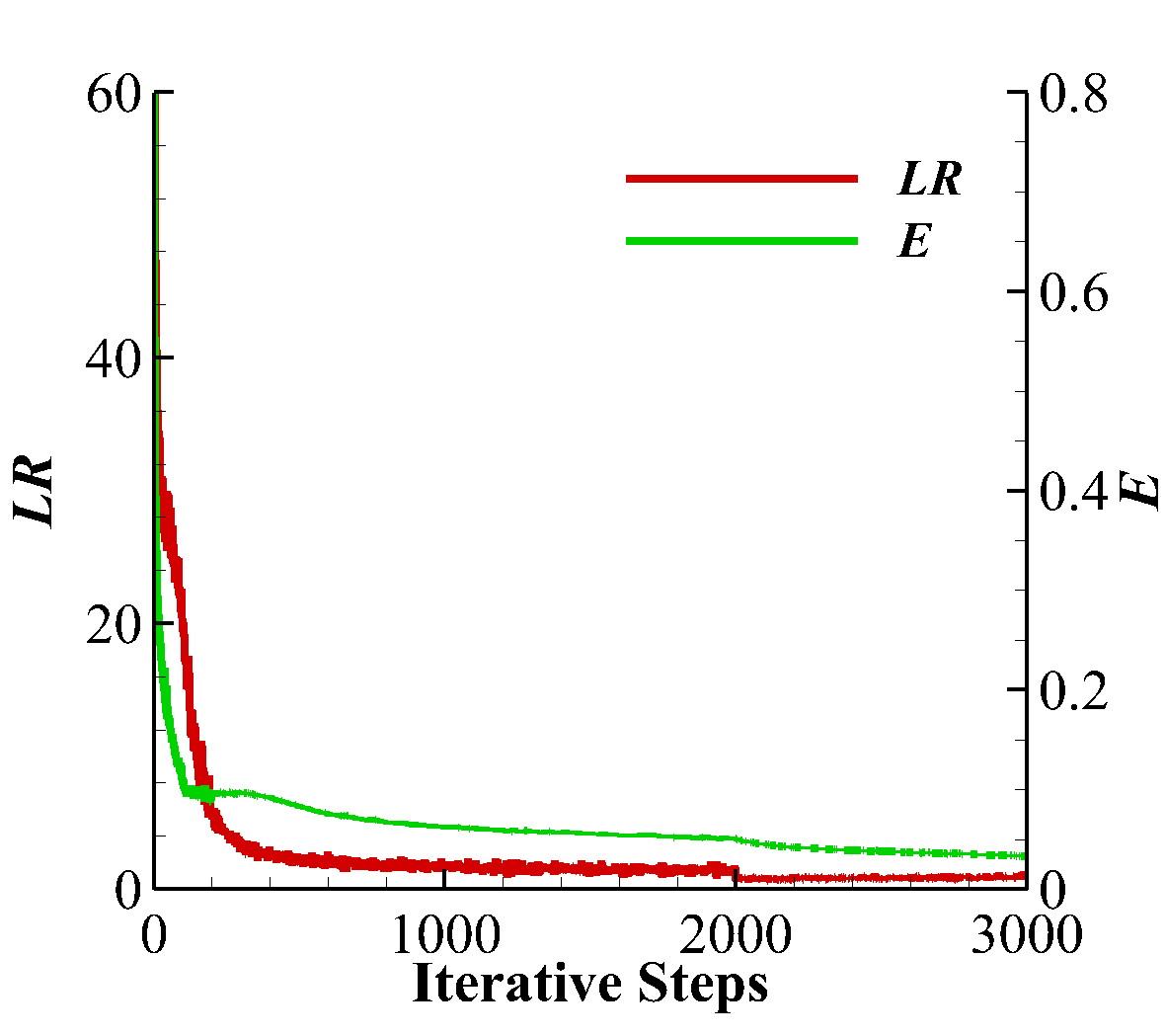}}
	\caption{Enhancements for physics-driven method: $LR$ and $E$ (threshold = 0.1). Except for $T_{\rm zero}$, all the reference targets are able to accelerate the learning notably.}
	\label{figs:14-Physics-driven Based Method}
\end{figure}

Due to the introduction of $E$ as one loss-term, the convergence speeds with the true and coarse targets are improved a lot.
As shown in Figure \ref{figs:14-Physics-driven Based Method},
compared with the steady decline in the physics-driven learning, $E$ of combined method drops more dramatically in the beginning.
We set the threshold $\mathcal{L}_{\rm{thr}}$ in Equation \ref{con:phybasedLoss} as 0.1. After $\mathcal{L}_{\rm{thr}}$ reached, only $LR$ is the loss term.
So $E$ changes to steady descent immediately while $LR$ still goes on rapid decline.
For $T_{\rm tar} =T_{\rm zero}$, $LR$ and $E$ have the similar trend at the very beginning. However, the overall error $E$ never reaches the threshold as the optimizer cannot find a way to reduce the gradient because of the incorrect reference target.

\begin{figure}[!h]
	\centering
	\subfloat[Physics-driven Method]{
		\includegraphics[width=0.33\textwidth]{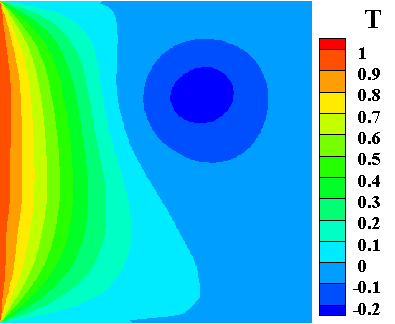}}
	\subfloat[$T_{\rm tar} =T_{\rm true} $]{
		\includegraphics[width=0.33\textwidth]{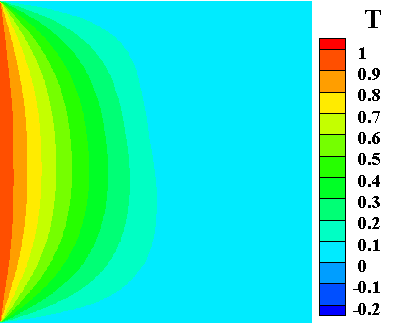}}
	\subfloat[$T_{\rm tar} =T_{\rm zero} $]{
		\includegraphics[width=0.33\textwidth]{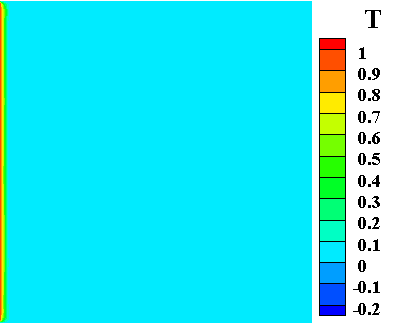}}\\
	\subfloat[$T_{\rm tar} =T_{\rm coarse} $]{
		\includegraphics[width=0.33\textwidth]{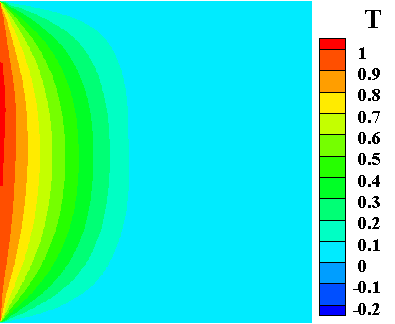}}
	\subfloat[$T_{\rm tar} =T_{\rm systemaitc} $]{
		\includegraphics[width=0.33\textwidth]{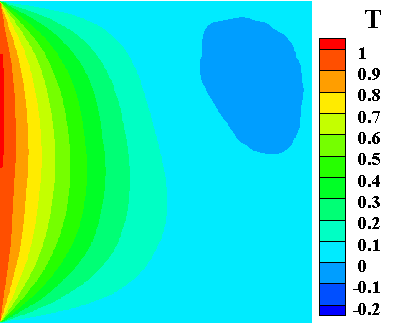}}
	\subfloat[$T_{\rm tar} =T_{\rm random} $]{
		\includegraphics[width=0.33\textwidth]{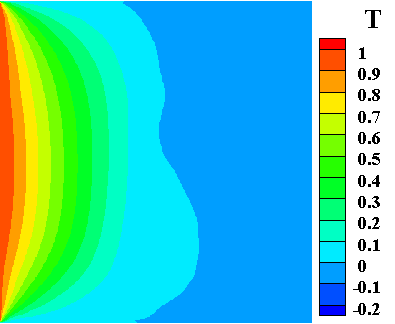}}
	\caption{Enhancements for physics-driven method: outputs of CNN (threshold = 0.1, iterative step = 1000). $T_{\rm zero}$ is not able to obtain the right solution. The other four reference targets are able to remedy the ``valley'' faster. }
	\label{figs:15-Physics-driven Based Method Field}
\end{figure}

Compared with the original physics-driven method, it is observed that the combined method with all the reference targets, except for zero profiles, are able to obtain a more accurate result by eliminating the large error spot quickly (as shown in Figure \ref{figs:15-Physics-driven Based Method Field}).
These results suggest that the requirement for appropriate reference targets are not very restrictive in practical applications.

To study the influence of the threshold $\mathcal{L}_{\rm{thr}}$, the mean costs of 5 independent trainings with different reference targets are summarized in Table \ref{tab:steps}. 
It is observed that the combined method improves physics-driven learning considerably. 
Compared with the true reference target, the other three coarse targets have a only slightly slower convergence speed.
With decreasing $\mathcal{L}_{\rm{thr}}=$0.15, 0.1 and 0.05, the acceleration obtained by using the true coarse target over the original physics-driven method are 22.7\%, 34.4\% and 49.0\% respectively. 
Similar behaviors have been obtained for the other two coarse references.
It is concluded that the smaller thresholds result in the bigger improvements. 
While in a real application, a too small threshold may result in the non-convergence as suggested by the zero profile reference. So one may choose a moderate threshold for safety.

\begin{table}[!htbp]
	\centering
	\caption{Costs of different methods. The cost is represented as the number of iteration steps when $E$ reaches $\mathcal{L}_{\rm{thr}}$ and the convergence absolute criteria $\mathcal{C}$ (0.01).}
	\begin{tabular}{cc|cc|cc|cc}
		\toprule
		\toprule
		\multicolumn{2}{c|}{\multirow{2}[4]{*}{Methods}} & \multicolumn{2}{p{6.71em}|}{\quad \ $\mathcal{L}_{\rm{thr}}=0.15$} & \multicolumn{2}{p{6.71em}|}{\quad \ $\mathcal{L}_{\rm{thr}}=0.1$} & \multicolumn{2}{p{6.71em}}{\quad \  $\mathcal{L}_{\rm{thr}}=0.05$} \\
		\cmidrule{3-8}    \multicolumn{2}{c|}{} & \multicolumn{1}{p{3.355em}}{\ \ $\mathcal{L}_{\rm{thr}}$} & \multicolumn{1}{p{3.355em}|}{\quad \ $\mathcal{C}$} & \multicolumn{1}{p{3.355em}}{\ \ $\mathcal{L}_{\rm{thr}}$} & \multicolumn{1}{p{3.355em}|}{\quad \ $\mathcal{C}$} & \multicolumn{1}{p{3.355em}}{\ \ $\mathcal{L}_{\rm{thr}}$} & \multicolumn{1}{p{3.355em}}{\quad \ $\mathcal{C}$} \\
		\midrule
		\multicolumn{2}{p{6.71em}|}{Physics-driven} & 1322  & 13494 & 2542  & 13494 & 6334  & 13494 \\
		\multicolumn{2}{p{6.71em}|}{\quad$T_{\rm true}$} & 51    & 9875  & 112   & 8904  & 138   & 6758 \\
		\multicolumn{2}{p{6.71em}|}{\quad$T_{\rm coarse}$} & 57    & 10428 & 109   & 8853  & 145   & 6876 \\
		\multicolumn{2}{p{6.71em}|}{\quad$T_{\rm systematic}$} & 48    & 10091 & 116   & 9025  & 140   & 6894 \\
		\multicolumn{2}{p{6.71em}|}{\quad$T_{\rm random}$} & 53    & 9917  & 105   & 8963  & 139   & 7054 \\
		\bottomrule
		\bottomrule
	\end{tabular}%
	\label{tab:steps}%
\end{table}%

\section{Conclusion}\label{sec:conclusion}
In this paper, we proposed a combined data-driven and physics-driven method to directly predict field solution of physics problems using deep CNN. 
The combined method simultaneously utilizes training data and physics law to drive the learning process. 
For the data-driven based method, besides accelerated convergence, the obtained local structure of solution achieves a better physics consistency. 
For the physics-driven based method, learning process can be accelerated considerably even with not very restrictive choices of reference targets, which is useful for practical application when a accurate reference is not available.
It is noteworthy that the related CNN architecture and heat conduction problem are generic and the combined method suits for other physics laws which can be expressed as PDEs. 
Further research will be carried out for predicting complex flow field with the present method.data

\section{Acknowledgement}

Hao Ma (No. 201703170250) and Yuxuan Zhang (No. 201804980021) are supported by China Scholarship Council when they conduct the work this paper represents. 
The authors thank the colleagues for the beneficial discussion, especially Chi Zhang and Nikolaos Perakis.

\clearpage


\bibliography{mybibfile}

\begin{thebibliography}{10}
\expandafter\ifx\csname url\endcsname\relax
  \def\url#1{\texttt{#1}}\fi
\expandafter\ifx\csname urlprefix\endcsname\relax\def\urlprefix{URL }\fi
\expandafter\ifx\csname href\endcsname\relax
  \def\href#1#2{#2} \def\path#1{#1}\fi

\bibitem{yan2018data}
W.~Yan, S.~Lin, O.~L. Kafka, Y.~Lian, C.~Yu, Z.~Liu, J.~Yan, S.~Wolff, H.~Wu,
  E.~Ndip-Agbor, et~al., Data-driven multi-scale multi-physics models to derive
  process--structure--property relationships for additive manufacturing,
  Computational Mechanics 61~(5) (2018) 521--541.

\bibitem{zhang2015machine}
Z.~J. Zhang, K.~Duraisamy, Machine learning methods for data-driven turbulence
  modeling, in: 22nd AIAA Computational Fluid Dynamics Conference, 2015, p.
  2460.

\bibitem{kirchdoerfer2016data}
T.~Kirchdoerfer, M.~Ortiz, Data-driven computational mechanics, Computer
  Methods in Applied Mechanics and Engineering 304 (2016) 81--101.

\bibitem{christelis2017physics}
V.~Christelis, A.~Mantoglou, Physics-based and data-driven surrogate models for
  pumping optimization of coastal aquifers, European Water 57 (2017) 481--488.

\bibitem{kim2017data}
M.~Kim, G.~Pons-Moll, S.~Pujades, S.~Bang, J.~Kim, M.~J. Black, S.-H. Lee,
  Data-driven physics for human soft tissue animation, ACM Transactions on
  Graphics (TOG) 36~(4) (2017) 54.

\bibitem{schmid2010dynamic}
P.~J. Schmid, Dynamic mode decomposition of numerical and experimental data,
  Journal of fluid mechanics 656 (2010) 5--28.

\bibitem{brunton2019machine}
S.~L. Brunton, B.~R. Noack, P.~Koumoutsakos, Machine learning for fluid
  mechanics, Annual Review of Fluid Mechanics 52 (2019).

\bibitem{anderson1995computational}
J.~D. Anderson, J.~Wendt, Computational fluid dynamics, Vol. 206, Springer,
  1995.

\bibitem{wang2017physics}
J.-X. Wang, J.-L. Wu, H.~Xiao, Physics-informed machine learning approach for
  reconstructing reynolds stress modeling discrepancies based on dns data,
  Physical Review Fluids 2~(3) (2017) 034603.

\bibitem{wu2018physics}
J.-L. Wu, H.~Xiao, E.~Paterson, Physics-informed machine learning approach for
  augmenting turbulence models: A comprehensive framework, Physical Review
  Fluids 3~(7) (2018) 074602.

\bibitem{goodfellow2014generative}
I.~Goodfellow, J.~Pouget-Abadie, M.~Mirza, B.~Xu, D.~Warde-Farley, S.~Ozair,
  A.~Courville, Y.~Bengio, Generative adversarial nets, in: Advances in neural
  information processing systems, 2014, pp. 2672--2680.

\bibitem{farimani2017deep}
A.~B. Farimani, J.~Gomes, V.~S. Pande, Deep learning the physics of transport
  phenomena, arXiv preprint arXiv:1709.02432 (2017).

\bibitem{duraisamy2019turbulence}
K.~Duraisamy, G.~Iaccarino, H.~Xiao, Turbulence modeling in the age of data,
  Annual Review of Fluid Mechanics 51 (2019) 357--377.

\bibitem{ling2016reynolds}
J.~Ling, A.~Kurzawski, J.~Templeton, Reynolds averaged turbulence modelling
  using deep neural networks with embedded invariance, Journal of Fluid
  Mechanics 807 (2016) 155--166.

\bibitem{wu2018deep}
H.~Wu, A.~Mardt, L.~Pasquali, F.~Noe, Deep generative markov state models, in:
  Advances in Neural Information Processing Systems, 2018, pp. 3975--3984.

\bibitem{bhatnagar2019prediction}
S.~Bhatnagar, Y.~Afshar, S.~Pan, K.~Duraisamy, S.~Kaushik, Prediction of
  aerodynamic flow fields using convolutional neural networks, Computational
  Mechanics (2019) 1--21.

\bibitem{jeong2015data}
S.~Jeong, B.~Solenthaler, M.~Pollefeys, M.~Gross, et~al., Data-driven fluid
  simulations using regression forests, ACM Transactions on Graphics (TOG)
  34~(6) (2015) 199.

\bibitem{tompson2017accelerating}
J.~Tompson, K.~Schlachter, P.~Sprechmann, K.~Perlin, Accelerating eulerian
  fluid simulation with convolutional networks, in: Proceedings of the 34th
  International Conference on Machine Learning-Volume 70, JMLR. org, 2017, pp.
  3424--3433.

\bibitem{thuerey2018well}
N.~Thuerey, K.~Weissenow, H.~Mehrotra, N.~Mainali, L.~Prantl, X.~Hu, Deep
  learning methods for reynolds-averaged navier-stokes simulations of airfoil
  flows, arXiv preprint arXiv:1810.08217 (2018).
\newblock \href {http://arxiv.org/abs/1810.08217v2}
  {\path{arXiv:1810.08217v2}}.

\bibitem{milani2018machine}
P.~M. Milani, J.~Ling, G.~Saez-Mischlich, J.~Bodart, J.~K. Eaton, A machine
  learning approach for determining the turbulent diffusivity in film cooling
  flows, Journal of Turbomachinery 140~(2) (2018) 021006.

\bibitem{lee2018data}
S.~Lee, D.~You, Data-driven prediction of unsteady flow fields over a circular
  cylinder using deep learning, arXiv preprint arXiv:1804.06076 (2018).

\bibitem{lee1990neural}
H.~Lee, I.~S. Kang, Neural algorithm for solving differential equations,
  Journal of Computational Physics 91~(1) (1990) 110--131.

\bibitem{lagaris1998artificial}
I.~E. Lagaris, A.~Likas, D.~I. Fotiadis, Artificial neural networks for solving
  ordinary and partial differential equations, IEEE transactions on neural
  networks 9~(5) (1998) 987--1000.

\bibitem{raissi2017physics}
M.~Raissi, P.~Perdikaris, G.~E. Karniadakis, Physics informed deep learning
  (part ii): Data-driven discovery of nonlinear partial differential equations,
  arXiv preprint arXiv:1711.10566 (2017).

\bibitem{raissi2017physicspart1}
M.~Raissi, P.~Perdikaris, G.~E. Karniadakis, Physics informed deep learning
  (part i): Data-driven solutions of nonlinear partial differential equations,
  arXiv preprint arXiv:1711.10561 (2017).

\bibitem{raissi2019physics}
M.~Raissi, P.~Perdikaris, G.~E. Karniadakis, Physics-informed neural networks:
  A deep learning framework for solving forward and inverse problems involving
  nonlinear partial differential equations, Journal of Computational Physics
  378 (2019) 686--707.

\bibitem{lu2019deepxde}
L.~Lu, X.~Meng, Z.~Mao, G.~E. Karniadakis, Deepxde: A deep learning library for
  solving differential equations, arXiv preprint arXiv:1907.04502 (2019).

\bibitem{sun2019surrogate}
L.~Sun, H.~Gao, S.~Pan, J.-X. Wang, Surrogate modeling for fluid flows based on
  physics-constrained deep learning without simulation data, arXiv preprint
  arXiv:1906.02382 (2019).

\bibitem{sharma2018weakly}
R.~Sharma, A.~B. Farimani, J.~Gomes, P.~Eastman, V.~Pande, Weakly-supervised
  deep learning of heat transport via physics informed loss, arXiv preprint
  arXiv:1807.11374 (2018).

\bibitem{zhu2019physics}
Y.~Zhu, N.~Zabaras, P.-S. Koutsourelakis, P.~Perdikaris, Physics-constrained
  deep learning for high-dimensional surrogate modeling and uncertainty
  quantification without labeled data, Journal of Computational Physics 394
  (2019) 56--81.

\bibitem{geneva2019modeling}
N.~Geneva, N.~Zabaras, Modeling the dynamics of pde systems with
  physics-constrained deep auto-regressive networks, arXiv preprint
  arXiv:1906.05747 (2019).

\bibitem{ronneberger2015u}
O.~Ronneberger, P.~Fischer, T.~Brox, U-net: Convolutional networks for
  biomedical image segmentation, in: International Conference on Medical image
  computing and computer-assisted intervention, Springer, 2015, pp. 234--241.

\bibitem{paszke2019pytorch}
A.~Paszke, S.~Gross, F.~Massa, A.~Lerer, J.~Bradbury, G.~Chanan, T.~Killeen,
  Z.~Lin, N.~Gimelshein, L.~Antiga, et~al., Pytorch: An imperative style,
  high-performance deep learning library, in: Advances in Neural Information
  Processing Systems, 2019, pp. 8024--8035.

\bibitem{kingma2014adam}
D.~P. Kingma, J.~Ba, Adam: A method for stochastic optimization, arXiv preprint
  arXiv:1412.6980 (2014).

\bibitem{jasak2007openfoam}
H.~Jasak, A.~Jemcov, Z.~Tukovic, et~al., Openfoam: A c++ library for complex
  physics simulations, in: International workshop on coupled methods in
  numerical dynamics, Vol. 1000, IUC Dubrovnik Croatia, 2007, pp. 1--20.

\bibitem{ruder2016overview}
S.~Ruder, An overview of gradient descent optimization algorithms, arXiv
  preprint arXiv:1609.04747 (2016).

\bibitem{panchapagesan2016multi}
S.~Panchapagesan, M.~Sun, A.~Khare, S.~Matsoukas, A.~Mandal, B.~Hoffmeister,
  S.~Vitaladevuni, Multi-task learning and weighted cross-entropy for dnn-based
  keyword spotting., in: Interspeech, Vol.~9, 2016, pp. 760--764.

\bibitem{redmon2016you}
J.~Redmon, S.~Divvala, R.~Girshick, A.~Farhadi, You only look once: Unified,
  real-time object detection, in: Proceedings of the IEEE conference on
  computer vision and pattern recognition, 2016, pp. 779--788.

\bibitem{phan2017dnn}
H.~Phan, M.~Krawczyk-Becker, T.~Gerkmann, A.~Mertins, Dnn and cnn with weighted
  and multi-task loss functions for audio event detection, arXiv preprint
  arXiv:1708.03211 (2017).

\end{thebibliography}

\end{document}